\documentclass[showpacs,amssymb,preprint,preprintnumbers 
,nofootinbib,superscriptaddress]{revtex4-2}

\usepackage{amsmath}
\usepackage{graphicx}
\graphicspath{{Figures/}}
\usepackage{latexsym}
\usepackage{amsfonts}
\usepackage{url,hyperref}
\usepackage{bm}
\usepackage{textcomp}
\usepackage[dvipsnames]{xcolor} 
\usepackage{bbm}
\usepackage{slashed}
\usepackage{epstopdf}
\usepackage{placeins}
\usepackage{array}
\usepackage{cancel}
\usepackage{verbatim}
\usepackage{caption}
\usepackage{subcaption}
\newcommand{\beq}{\begin{equation}}
\newcommand{\eeq}{\end{equation}}
\def\bea{\begin{eqnarray}}
\def\eea{\end{eqnarray}}

\hypersetup{colorlinks=true,linkcolor=blue,citecolor=magenta,filecolor=magenta,
  urlcolor=blue}
\renewcommand{\theequation}{\arabic{section}.\arabic{equation}}

\DeclareUnicodeCharacter{2212}{-} 

\usepackage{etoolbox}

\makeatletter
\patchcmd{\frontmatter@abstract@produce}
  {\vskip200\p@\@plus1fil
   \penalty-200\relax
   \vskip-200\p@\@plus-1fil}
  {}
  {}
  {}
\makeatother

\newcolumntype{C}{>{$}c<{$}} 
\newcolumntype{L}{>{$}l<{$}} 

\begin{document}

\title{Quasinormal modes for integer and half-integer spins\\within the large angular  momentum limit}

\author{Chun-Hung Chen}
\email{chun-hungc@nu.ac.th}
\affiliation{The Institute for Fundamental Study, \\
Naresuan University, Phitsanulok 65000, Thailand.}

\author{Hing-Tong Cho}
\email{htcho@mail.tku.edu.tw}
\affiliation{Department of Physics, Tamkang University,\\
Tamsui District, New Taipei City, Taiwan 25137.}

\author{Anna Chrysostomou}
\email{annachrys97@gmail.com}
\affiliation{Department of Physics, University of Johannesburg, \\
PO Box 524, Auckland Park 2006, South Africa.}

\author{Alan S. Cornell}
\email{acornell@uj.ac.za}
\affiliation{Department of Physics, University of Johannesburg, \\
PO Box 524, Auckland Park 2006, South Africa.}


\begin{abstract}
While independent observations have been made regarding the behaviour of effective quasinormal mode (QNM) potentials within the large angular momentum limit, we demonstrate analytically here that a uniform expression emerges for non-rotating, higher-dimensional, and spherically-symmetric black holes (BHs) in this regime for fields of integer and half-integer spin in asymptotically flat and dS BH contexts; a second uniform expression arises for these QNM potentials in AdS BH spacetimes. We then proceed with a numerical analysis based on the multipolar expansion method recently proposed by Dolan and Ottewill to determine the behaviour of quasinormal frequencies (QNF) for varying BH parameters in the eikonal limit. We perform a complete study of Dolan and Ottewill's method for perturbations of spin $s \in \{0,1/2,1,3/2,2 \}$ in 4D Schwarzschild, Reissner-Nordstr{\"o}m, and Schwarzschild de Sitter spacetimes, clarifying expressions and presenting expansions and results to higher orders $(\mathcal{O}(L^{-6}))$ than many of those presented in the literature $(\sim \mathcal{O}(L^{-2}))$. We find good agreement with known results of QNFs for low-lying modes; in the large-$\ell$ regime, our results are highly consistent with those of Konoplya's 6th-order WKB method. We confirm a universality in the trends of physical features recorded in the literature for the low-lying QNFs (that the real part grows indefinitely, the imaginary tends to a constant as $\ell \rightarrow \infty$, etc.) as we approach large values of $\ell$ within these spacetimes, and explore the consequent interplay between BH parameters and QNFs in the eikonal limit.  

\end{abstract}

{\pacs{04.20.−q, 04.40.−b, 04.50.Gh }}

\date{\today}

\maketitle

%
%
\section{\label{sec:intro}Introduction}
\par An isolated black hole (BH) is a simple object: mathematically, it represents the vacuum solution of Einstein's field equations \cite{refSchwarzschild}; astrophysically, its full description can be attained with the three parameters of Arnowitt-Deser-Misner (ADM) mass ($M$), charge ($Q$), and angular momentum ($a$) when in equilibrium \cite{refHeusler}. Despite this apparent simplicity, BHs provide a medium through which we might explore gauge-gravity duality (AdS/CFT) and physics beyond the Standard Model (BSM) that range from supersymmetry (SUSY) models to theories of quantum gravity \cite{refBertiCardoso, refKonoplyaZhidenkoReview}. With the advent of gravitational-wave (GW) astronomy \cite{refLIGO}, we can now exploit perturbed BHs as GW sources, the data from which may be utilised in constraining BSM conjectures, such as SUSY models, extra-dimensional frameworks, etc. \cite{Yu2019} and test long-standing hypotheses like the ``no-hair" conjecture \cite{refNoHair}.

\par With this in mind, we concern ourselves with the quasinormal mode (QNM), which is a fundamental feature of the damped ``ringdown" phase through which a perturbed system passes as it returns to equilibrium. The corresponding quasinormal frequencies (QNFs) are complex, such that they may be decomposed into a real and an imaginary component,
\begin{equation}
\omega = \omega_R - i \omega_I \;, \hspace{1cm} \omega_R, \omega_I \in \mathbb{R} \;,
\end{equation}
where the real part represents the physical oscillation frequency and the imaginary part expresses the damping (see Refs. \cite{refBertiCardoso,refKonoplyaZhidenkoReview} for comprehensive reviews).

\par For spherically-symmetric BHs, the QNM wavefunction can be separated into its radial and angular components, where the latter can be expressed fully through spherical harmonic decompositions. As first illustrated by Regge and Wheeler \cite{refRW}, extended by Zerilli \cite{refZerilli,ZerMon1} and Moncrief \cite{ZerMon2,ZerMon3,ZerMon4}, then generalised fully by Ishibashi and Kodama \cite{refIKschwarz1,refIKschwarz2,refIKrn,refIKchap6}, the radial QNM behaviour can be described in the weak-field limit by a second-order ordinary differential equation,
\begin{equation} \label{eq:ode}
\frac{d^2 \psi}{dr_*^2} + \left[ \omega^2 - V(r) \right] \psi (r_*) = 0 \;.
\end{equation}
This requires the introduction of a ``tortoise coordinate",
\begin{equation}
\frac{d}{dr_*} = f(r) \frac{d}{dr} \;,
\end{equation}
where $f(r)$ refers to the metric function for a spherically-symmetric BH through the expression,
\begin{equation} \label{eq:f(r)}
f(r) = 1 - \frac{2 \mu}{r^{d-3}} + \frac{\theta^2}{r^{2(d-3)}} - \lambda r^2 \;.
\end{equation}
Here, $\mu$, $\theta$, and $\lambda$ parametrise $M$, $Q$, and the cosmological constant $\Lambda$:
\[ \mu = \frac{8 \pi G_d}{(d-2) \; \Omega_{d-2}} \;M \;, \hspace{1cm} \theta^2 = \frac{8 \pi G_d}{(d-2) (d-3)} \; Q^2 ,\hspace{0.6cm} \text{and} \hspace{0.6cm} \lambda=\frac{2\Lambda}{(d-1)(d-2)} \]
for the gravitational constant in $d$-dimensional space $G_d$, and the area of a unit $(d-2)$-sphere $\Omega_{d-2}$. Minkowski, de Sitter (dS) and anti-de Sitter (AdS) spacetimes are denoted by $\lambda = 0$, $\lambda > 0$, and $\lambda <0$, respectively \cite{refIKschwarz1,refIKschwarz2,refIKrn}.

\par To capture its behaviour fully, the QNM wavefunction is associated with multiple parameters required for the spherical decompositions of the angular component and the description of the radial component. Of particular interest to us are those upon which the radial function depends: $s$, the spin of the oscillating field; $\ell$, the angular momentum (multipolar) number describing the angular dependence of the wave and serving as the eigenvalues for the spherical harmonics; $n$, the overtone number, which labels the QNMs by a monotonically increasing value of $\ell$ 
 \cite{refBertiCardoso,refKonoplyaZhidenkoReview}.

\par Much focus in the literature has been placed on the $n \rightarrow \infty$ limit, initiated by a conjectured link between quantum gravity and asymptotic QNF expressions \cite{Hod1998}. Interest in the the large $\ell$ limit, however, has been invested primarily within the context of mathematical physics. This is a consequence of the limitations of the numerical methods by which QNM calculations are often addressed. Well-established procedures such as the P{\"o}schl-Teller (PT) approximation method suggested in Ref. \cite{PoschTellerMethod} and the modified Wentzel-Kramers-Brillouin (WKB) approximation developed in Refs. \cite{refBHWKB0,refBHWKB1,refBHWKB2,refBHWKB3,refBHWKB4,Konoplya2003} $-$ as well as more recently constructed methods like the AIM \cite{refAIM_OG}, the improved AIM \cite{refAIM}, and the improved semianalytic approach \cite{Matyjasek2017,Matyjasek2019} $-$ are most accurate in the eikonal limit, and have been shown to break down for large overtones (see Refs. \cite{refKonoplyaZhidenkoReview,refZhidenko2004,Konoplya2019} and references therein). As such, subjecting the QNM potential to the $\ell \rightarrow \infty$ limit is required in order to extract the analytic expression needed to compute QNFs within these methods, and may serve as a means by which to assess the validity of various methods in QNM contexts (in a manner much like Ref. \cite{Konoplya2019}). Whether this can be exploited in the establishment of a mathematical and even machine-learning algorithm for QNF calculations remains to be seen, and is reserved for a future work.

\par Beyond its practical use in facilitating QNF calculations, the application of the large multipolar limit may offer physical insights into BH systems. This can be inferred from studies such as Ref. \cite{refGoebel1972}, in which the QNF spectrum for the gravitational field perturbations of a 4D Schwarzschild BH was found to be of the form
\begin{equation} \label{eq:Lyapunov}
\omega_{n, \ell \rightarrow \infty} = \Omega \left( \ell + \frac{1}{2} \right) - i \overline{\Lambda} \left( n + \frac{1}{2} \right) + \mathcal{O}(\ell^{-1}) \;, \hspace{1cm} \Omega = \overline{\Lambda} = \frac{1}{ \sqrt{27} M } \;,
\end{equation}
for large $\ell$ and fixed $n$. $\Omega$ here refers to the orbital frequency of the $r_{orb}=3M$ photon sphere and $\overline{\Lambda}$ is the Lyapunov exponent describing the decay time scale. This behaviour in the large multipolar limit was echoed in Ref. \cite{refCardosoLyapunov} at leading order, and found to apply also for scalar and electromagnetic perturbing fields by Ferrari and Mashhoon \cite{refFerrariMashhoon}. This was confirmed in Ref. \cite{refPanJing} via Leaver's CFM and Ref. \cite{refShuShen} via Konoplya's 6th-order WKB method \cite{Konoplya2003}.

\par The relationship between the Dirac QNFs and the multipolar number, first studied in Ref. \cite{refDirac03}, exhibited this same behaviour: $\mathbb{R}e \{ \omega \}$ was found to increase with $\ell$, whereas the magnitude of $\mathbb{I}m \{ \omega \}$ increased with $n$. This was confirmed in Ref. \cite{refShuShen}.

\par This dependence of QNFs on $\ell$ has been noted also in more complicated spacetimes. One such example of this is the numerical analysis by Zhidenko \cite{refZhidenko2004} for the QNFs of a 4D SdS BH spacetime. For arbitrary spins, the QNF within the large multipolar limit yielded eq. (\ref{eq:Lyapunov}), albeit with
\begin{equation} \label{eq:LyapunovSdS}
\Omega = \overline{\Lambda} = \frac{\sqrt{1-9M^2 \Lambda}}{\sqrt{27}M} \;.
\end{equation}
\noindent In Ref. \cite{refWangLinMolina}, Wang, Lin, and Molina came to similar conclusions when studying scalar perturbations within Reissner-Nordstr{\"o}m AdS spacetimes for $\ell \in [1,10]$: the imaginary part of the QNF was shown to decrease, while the real increased almost linearly, with larger multipolar numbers. These same relationships between the angular momentum and the components of the QNF were observed in Ref. \cite{refWangHerdeiroJing} for Schwarzschild AdS spacetimes perturbed by a spin-1/2 field.

\par Next, we note that the study of the QNMs in reference to the multipolar number has already yielded tangible outcomes. On the basis of geometric arguments derived from the form of eq. ({\ref{eq:Lyapunov}}), Dolan and Ottewill in Ref. \cite{refDolanOttewill2009} established a novel ansatz by which eq. (\ref{eq:ode}) may be solved for QNMs. Their numerical method involves expansion in inverse powers of $L=\ell + 1/2$. This allows for the analysis of higher-order asymptotics and the production of accurate results in asymptotically flat and dS spherically-symmetric spacetimes \cite{refBertiCardoso}, as shown for the integer-spin perturbations discussed in Ref. \cite{refDolanOttewill2009}. To date, Dolan and Ottewill's method has been applied to a number of BH contexts (see Refs. \cite{refDecanini2011,refFernandoCorrea,refLiMaLin2013,refLiLinYang}), but often only to low orders of $\mathcal{O}(L^{-k})$. Furthermore, by virtue of the fact that QNMs can be defined as the poles of a Green's function, this geometric interpretation of QNMs as perturbations of unstable null geodesics can be extended to investigations of the self-force of compact objects \cite{refDolanOttewill2013} and the analysis of the singular structure of the Green's function \cite{refDolanOttewill2011}. The latter is especially useful due to the relevance of Green's functions in AdS/CFT considerations, and their nature as propagators for the Klein-Gordon equation \cite{refBertiCardoso}.

\par In this paper, our main focus is to understand the physical implications of imposing the large angular momentum limit on QNFs within spherically-symmetric BH spacetimes. In section \ref{sec:potentials}, we collate known effective potentials for integer and half-integer QNMs of non-rotating, spherically-symmetric BH spacetimes and demonstrate how their consistent behaviour in the large-$\ell$ regime within BH spacetimes of $\lambda \geq 0$ differs from that of $\lambda < 0$. We then proceed to a numerical study of the large-$\ell$ limit in section \ref{sec:num}: we perform a complete study of QNFs for spin $s \in \{ 0,1/2,1,3/2,2 \}$ within 4D Schwarzschild, Reissner-Nordstr{\"o}m (RN), and Schwarzschild de Sitter (SdS) BH spacetimes for large multipolar numbers, where we confine our numerical work to 4D BHs to produce astrophysically-relevant results. We provide explicit descriptions of the necessary ansatz, inverse multipolar expansions, and associated parameters for each context and carry the method to orders of $\mathcal{O}(L^{-6})$ in almost all cases. This marks an improvement on extant results for some integer-spin cases and a production of new results for integer-spin QNFs in RN and SdS BH spacetimes, as well as new results for half-integer QNFs in all BH spacetimes studied. To our knowledge, this is the first time this method has been applied to QNFs of spin-3/2. 

\par We observe that Dolan and Ottewill's method is well-suited for our purpose due to its physically-motivated foundations and the ease with which it lends itself to computations within the eikonal limit. When employing low- and high-lying results to validate the method against existing techniques, we find consistency with the literature in the low-$\ell$ regime (see appendix \ref{app}) and excellent agreement with Konoplya's 6th-order WKB method \cite{Konoplya2003} in the large-$\ell$ regime. From the QNF expressions we obtain for large values of $\ell$, we deduce the effect of $\theta$ and $\lambda$ on QNFs in the eikonal limit. These and other observations are collected and commented upon in section \ref{sec:conc}.
\vfill

%
%
\section{\label{sec:potentials}Potentials in the large multipolar limit}

%
\subsection{\label{subsec:integer}Integer spin perturbations}

\par The well-known expressions for the effective potential of integer-spin QNMs in 4D Schwarzschild spacetimes can be written concisely in the ``Regge-Wheeler" form \cite{refRW,refBertiCardoso,refKonoplyaZhidenkoReview},
\begin{equation} \label{eq:RWconcise}
V_{eff}(r) = \frac{f(r)}{r^2} \left[ \ell (\ell+1) + \frac{2 \mu (1 - s^2)}{r} \right] \;,
\end{equation}
\noindent where
\[
s= \begin{cases}
0 \;, \; \;\text{scalar perturbations} & \Rightarrow (1-s^2) = \; \; 1 \\
1 \;, \; \; \text{electromagnetic perturbations: scalar/vector mode } & \Rightarrow  (1-s^2) = \; \; 0 \\
2 \;, \; \; \text{gravitational perturbations: vector mode} & \Rightarrow (1-s^2) = -3  \;.
\end{cases}
\]
This expression serves as a description of the radial dependence of a spin-$s$ perturbation of the spacetime, with which the angular momentum of $\ell$ is associated \cite{refBertiCardoso, refKonoplyaZhidenkoReview}. Furthermore, it demonstrates explicitly that irrespective of the spin, the dependence of the effective potential on the multipolar number adheres to a distinct form proportional to $\ell(\ell+1)f(r)/r^2$ within the asymptotically flat, four-dimensional Schwarzschild BH spacetime.

\par Within a RN BH spacetime, these effective QNM potentials can be similarly condensed into the ``Moncrief-Zerilli" form \cite{ZerMon1,ZerMon2,ZerMon3,ZerMon4,refOnozawa1996},
\begin{equation} \label{eq:ZMconcise}
V_{eff}(r) = \frac{f(r)}{r^2} \left[ \left( \ell (\ell + 1 \right) - \frac{q_s}{r} + \frac{\theta^2  p_s }{r^2} \right] \;,
\end{equation}
\noindent where for the scalar, electromagnetic (scalar/vector mode), and gravitational (vector mode) perturbations,
\[
s= \begin{cases}
0 \;, & p_0 = 2 \;, \hspace{0.4cm}  q_0 = -2 \\
1 \;, & p_1 = 4 \;, \hspace{0.4cm}  q_1 = 3\mu - \sqrt{9 \mu^2 + 4 \theta^2 \left( \ell (\ell + 1 ) - 2 \right)  \;}  \\
2 \;, &  p_2=4 \;, \hspace{0.4cm}  q_2 = 3\mu + \sqrt{9 \mu^2 + 4 \theta^2 \left( \ell (\ell + 1 ) - 2 \right) \;} \;.
\end{cases}
\]

\noindent While the $\ell (\ell + 1)f(r)/r^2$ dependence characteristic of the 4D non-rotating, spherically-symmetric BH spacetime remains, the relationship between effective potential and angular momentum in the cases of the electromagnetic and gravitational field is complicated by the square root term, as shall be discussed more fully in section \ref{subsec:RN}.

\par Note that we specify the spin-2 and spin-1 fields as the ``vector mode". To describe the gravitational field perturbations fully, one requires the vector-type (i.e. Regge-Wheeler/odd-parity/axial) as well as the scalar-type (i.e. Zerilli/even-parity/polar) mode \cite{refZerilli}. These have been shown to be isospectral \cite{refIsospectralRWZ}. Upon entering into discussions within higher-dimensional spacetimes, an additional tensor-type mode must also be included. Isospectrality no longer holds in this context \cite{refIKchap6}. For electromagnetic perturbations, the full description in higher dimensional spacetimes is attained when a scalar-type mode is also introduced.

\par Furthermore, if spacetime curvature is also to be considered, the effective potential gains a term $\thicksim -\lambda r^2$. In a manner analogous to eq. ({\ref{eq:RWconcise}}), we may summarise these for a $d$-dimensional Schwarzschild BH in (A)dS spacetimes as
\begin{equation} \label{eq:RWconciseHighD}
V_{eff}(r) = \frac{f(r)}{r^2} \left[ \ell (\ell+d-3) + \frac{(d-2)(d-4)}{4} - \frac{K}{4}\; \lambda r^2 + \frac{P}{2r^{d-3}} \; \mu \right] \;,
\end{equation}
where the forms of $K$ and $P$ are associated with the perturbation of interest, as summarised in Table \ref{table1}.

\begin{table}[t]
\centering
\caption{\textit{The values of $K$ and $P$ for the effective QNM potential of non-rotating, spherically-symmetric, higher-dimensional S(A)dS BH spacetimes.}}
\begin{tabular}{|l|c|c|}
\hline
\hspace{0.7cm} perturbation type & $K$ & $P$ \\
\hline
scalar \cite{refBertiCardoso,refCardosoDiasLemos}  & $d(d-2)$ & ${\color{white}{-3}}(d-2)^2$ \\
\hline
electromagnetic: scalar \cite{Lopez-Ortega2006} & $ {\color{white}{-}}(d-2)(d-4)$ & ${\color{white}{-}} d(d-4)$ \\
electromagnetic: vector \cite{Lopez-Ortega2006}  & $(d-4)(d-6)$ & $-(d-4)(3d-8)$ \\
\hline
gravitational: vector \cite{refIKschwarz1} & $ (d-2)(d-4)$ & $-3(d-2)^2$ \\
gravitational: tensor \cite{refIKschwarz1}  & $d(d-2)$ & ${\color{white}{-3}} (d-2)^2$ \\
\hline
\end{tabular}
\label{table1}
\end{table}

\par The form of the potentials remains similar for the higher-dimensional RN BHs, albeit with the additional terms required to account for the BH charge \cite{refBertiKokkotas,refIKrn}. In the cases of electromagnetic perturbations in RN BHs, and gravitational perturbations in all spherically-symmetric BHs, the scalar-type mode is a complicated expression that veers from the usual form exhibited by tensor- and vector-type modes (see Refs. \cite{refBertiKokkotas} and \cite{refIKschwarz1,refIKschwarz2,refIKrn}, respectively).

\par Upon reviewing these effective integer-spin QNM potentials within static and spherically-symmetric BH spacetimes $-$ charged or neutral, and exclusive of a cosmological constant $-$ we observe that the application of the eikonal limit yields a common form of
\begin{equation} \label{eq:bigL}
V_{eff} \bigg \vert_{\ell \rightarrow \infty} \approx \frac{f(r)}{r^2} \; \ell^2
\end{equation}
within Minkowski spacetimes. This is in keeping with  eq. (41) of Ref. \cite{refCardosoLyapunov} for the massless Klein-Gordon potential subjected to $\ell \rightarrow \infty$, which the authors justify as ``universal" for scalar, electromagnetic, and gravitational perturbations based on the uniformity of field behaviour reported in Ref. \cite{refIKrn} for these spacetimes.

\par With the inclusion of the cosmological constant comes the need to consider the asymptotic behaviour of $r$. For dS spacetimes, there exists a cosmological horizon towards which $r$ tends when its asymptotic behaviour is considered. As such, $r$ approaches a constant and finite value, and therefore does not interfere with the outcome of $V_{eff}$ under the influence of the large angular momentum limit; the ``uniform" potential of eq. ({\ref{eq:bigL}}) is still obtained for $\ell \rightarrow \infty$ within spacetimes of $\lambda >0$ for the fields mentioned above. Within AdS spacetimes, however, infinite limits must be considered carefully, as the nature of the boundary conditions and their effects on the behaviour of the potentials can be significant (demonstrated explicitly in Ref. \cite{refnewQNMs2020}). In this large-$\ell$ limit, the potential behaves as
\[ V_{eff} \big \vert_{r \rightarrow \infty} \propto r^2  \;, \]
\noindent which prevents us from dismissing the $\thicksim -\lambda r^2$ term as we ordinarily would. Consequently, eq. ({\ref{eq:bigL}}) becomes
\begin{equation} \label{eq:bigLAdS}
\frac{f(r)}{r^2} \; \ell^2  \; \; \rightarrow constant
\end{equation}
in the large $\ell$ limit for the integer-spin fields discussed here.

\par In the subsection that follows, we assess whether these expressions hold for the effective QNM potentials associated with perturbing fields of half-integer spin.

%
\subsection{\label{subsec:halfinteger}Half-integer spin perturbations}

\noindent For non-rotating and spherically-symmetric BH spacetimes, the effective Dirac \cite{refDirac03,refDirac07,refChakrabarti,refGiammatteo,refJing}
and Rarita-Schwinger \cite{refRS15,refRSSchwarz16,refRSRN18,refRSSAdS19} potentials have a supersymmetric form,
\begin{equation} \label{eq:SUSYpot}
V_{1,2} = \pm \frac{dW}{dr_*} + W^2 \;,
\end{equation}
where $d/dr_* = f(r) d/dr $ unless otherwise stated. $V_1$ and $V_2$ are isospectral supersymmetric partners \cite{refSUSYpot}. As shall be shown, the potentials are further categorised within the spin-3/2 framework into transverse traceless (TT) and non-transverse traceless (non-TT) eigenmodes in order to describe the perturbations of $d$-dimensional, spherically-symmetric BHs fully, according to the gauge-invariant formalism constructed in Refs. \cite{refRS15,refRSSchwarz16,refRSRN18,refRSSAdS19}. Since TT modes emerge only for $d \geq 5$, their study is confined to this analytical section.

\subsubsection*{Spin-1/2 perturbations}

\hspace{0.3cm} We first consider the higher-dimensional Schwarzschild BH as investigated in Ref. \cite{refDirac07}, characterised here by $f(r) =  1 - 2 \mu /r^{d-3} \;,$ where $\kappa = \ell + (d-2)/2$ for $\ell = 0,1,2,...$ is the spinor eigenvalue on the $(d-2)$-sphere. We calculate $V_1$ explicitly here by substituting $W=\sqrt{f(r)}\kappa/r$ and $f(r)$:
\begin{eqnarray}
V_1(r) &=& f(r) \frac{d}{dr} \left[ \sqrt{f(r)} \frac{\kappa}{r} \right] +  \left[f(r) \frac{\kappa^2}{r^2} \right] \nonumber \\
&=& \left[ \frac{r^{d-3}-2 \mu}{r^{d-3}} \right] \; \frac{d}{dr} \left[ \sqrt{\frac{r^{d-3}-2\mu}{r^{d-3}}} \; \frac{\kappa}{r} \right] + \; \; \left[\frac{r^{d-3}-2 \mu}{r^{d-3}} \right] \frac{\kappa^2}{r^2} \;. \label{eq:halfV1}
\end{eqnarray}
With the introduction of the definition, $ \Delta = r^{d-3} \left( r^{d-3}-2 \mu \right)$, we may simplify eq. (\ref{eq:halfV1}) to obtain
\begin{equation} \label{eq:halfV1final}
V_1 = \frac{\kappa \Delta^{1/2}}{r^{2(d-2)}} \left(\kappa \Delta^{1/2} -r^{d-3} + \left(\frac{d-1}{2} \right) 2 \mu \right) \;.
\end{equation}
This same expression emerges for $V_2$, as expected. We may then subject eq. ({\ref{eq:halfV1final}}) to $\kappa \rightarrow \infty$. This preserves only the terms in which $\kappa^2$ can be found, as these dominate. Thus, we obtain
\begin{equation}  \label{eq:bigkappa}
V_{1,2} \bigg \vert_{\kappa \rightarrow \infty} \approx \frac{\kappa^2 r^{d-3} \left(r^{d-3}-2 \mu \right)}{r^{2(d-2)}} = \kappa^2\frac{ f(r)}{r^2} \;.
\end{equation}

\par As such, we find that the massless Dirac field yields the same expression as that of eq. ({\ref{eq:bigL}}) within the large angular momentum limit. That this holds also for the $d$-dimensional RN BH was demonstrated in Ref. \cite{refChakrabarti}, following the method outlined in Ref. \cite{refDirac07}.

\par From investigations into the effective potentials of Dirac QNMs in 4D Schwarzschild \cite{refGiammatteo} and RN \cite{refJing,refWangHerdeiroJing} BH spacetimes inclusive of a cosmological constant, the application of the large $\ell$ limit to the effective potential once again produces the ``uniform" potential of eq. ({\ref{eq:bigL}}). However, as discussed for the integer spin cases, this result is applicable only for Minkowski and dS spacetimes; within AdS spacetimes, the spin-1/2 QNMs of Schwarzschild and RN BHs reduce to eq. ({\ref{eq:bigLAdS}}) when $\ell \rightarrow \infty,$ due to the influence of the asymptotic behaviour of $r$.

\subsubsection*{Spin-3/2 perturbations}

\noindent Let us begin with the Schwarzschild-Tangherlini BH studied in Ref. \cite{refRSSchwarz16}. For the TT eigenmodes in $d\geq 5$, the potentials for the spinor-vectors retain their supersymmetric form, with a superpotential defined as
\begin{equation}
{\mathbb{V}}_{1,2} = \pm \frac{d}{dr_*} \mathbb{W} + \mathbb{W}^2 \;, \hspace{1cm}
\mathbb{W} = \frac{\sqrt{f(r)}}{r} \; \zeta \;,
\end{equation}
for $\zeta = j + (d-3)/2 $ and $j= \ell + 1/2$ \cite{refRSSchwarz16}. Here, $\zeta$ is the spinor-vector eigenvalue on the $(d-2)$-sphere and $j =3/2,5/2,...$ . Since the metric function remains $f(r) =  1 - 2 \mu /r^{d-3} \;,$ an application of the large-$\ell$ limit naturally yields eq. (\ref{eq:bigkappa}), as in the Dirac case. This suggests that the TT eigenmode for the spin-3/2 field on the $N$-sphere is equivalent to the higher-dimensional spinor field. Since the premise of the spin-3/2 framework constructed in Ref. \cite{refRSSchwarz16} includes a convolution of spin-1/2 and spin-1 fields, this shared form is sensible. 
\par Though they retain their supersymmetric form, the potentials presented in Ref. \cite{refRSSchwarz16} for the non-TT eigenmodes are fairly complicated, with a superpotential expressed as
\begin{equation} \label{eq:WnonTTRSschwarz}
W = \frac{\sqrt{f(r)}}{r} \kappa \; \left[ \frac{\kappa^2 - \frac{(d-2)^2}{4} \left(1 + \frac{d-4}{d-2} \frac{2 \mu}{r^{d-3}} \right)}{\kappa^2 - \frac{(d-2)^2}{4} \left(1 - \frac{2 \mu}{r^{d-3}} \right)} \right] \;.
\end{equation}
\par However, if we insert the superpotential into eq. (\ref{eq:SUSYpot}) and extract only the terms dependent on the multipolar number, we obtain
\begin{eqnarray}
V_{1,2} &\thicksim& \frac{X\kappa \sqrt{f(r)}}{r^2(X+Y)^2} \left[ X \kappa \sqrt{f(r)} \right] + \frac{\kappa \sqrt{f(r)} Y^2}{r^2(X+Y)^2} \left[ \kappa \sqrt{f(r)} \left(\frac{d-4}{d-2} \right)^2\right] \notag \\
&& \hspace{3.5cm} + \frac{X \kappa \sqrt{f(r)}}{r^2(X+Y)^2} \left[ \pm 2(d-4)Y \right] \;.  \label{eq:RSScwarznontt}
\end{eqnarray}
\noindent As before, $\kappa = j + (d-3)/2$ is the spinor eigenvalue on the $(d-2)$-sphere for $j = 3/2,5/2,...$. We also define  
\[
X= \left(\frac{2}{d-2} \right)^2 \left(j - \frac{1}{2} \right) \left(j + \frac{2d-5}{2} \right)  \hspace{0.7cm} \text{and} \hspace{0.7cm} Y = \frac{2 \mu }{r^{d-3}}  \;.
\]
\noindent If we then isolate the $\mathcal{O}(\kappa^2)$  terms from eq. ({\ref{eq:RSScwarznontt}}) and simplify appropriately, we find that
\begin{equation}
V_{1,2} \bigg \vert_{\kappa \rightarrow \infty}
 \approx  \frac{\kappa^2 f(r)}{r^2(X+Y)^2} \left[ X^2 + Y^2 \left(\frac{d-4}{d-2} \right)^2\right] \approx  \frac{\kappa^2 (r^{d-3} - 2\mu)}{r^{d-1}} \;.
\end{equation}
Since we consider the $(d-4)/(d-2)$ term to be negligible for $\ell \rightarrow \infty$ and $d>4$, we observe that the non-TT potentials reduce to the same expression within the large angular momentum limit as their TT counterpart. This again yields the ``uniform" potential of eq. ({\ref{eq:bigL}}) for the large multipolar limit.

\par We may now turn to Ref. \cite{refRSRN18}, where a study of the spin-3/2 perturbations of higher-dimensional RN BHs yielded TT eigenmodes for which the potentials $\mathbb{V}_{1,2}$ were defined in terms of
\begin{equation} \label{eq:RSRNtermsC}
\mathbb{W} = \frac{\sqrt{f(r)}}{r} \left( \zeta - C \right) \;, \hspace{1cm} C = \frac{d-2}{2} \; \frac{\theta}{r^{d-3}} \;,
\end{equation}
\noindent with the spinor-vector eigenvalue given by $\zeta = j + (d-3)/2$ for $j = 3/2, 5/2, 7/2,...$. Upon substituting this into eq. ({\ref{eq:SUSYpot}}) and subjecting the expression to the $\zeta \rightarrow \infty$ limit, the consequent expression for the large multipolar limit is precisely that shown in eq. ({\ref{eq:bigL}}).

\par As we might expect, the non-TT potentials for the higher-dimensional RN BH prove more complicated. However, if define $\kappa = j + (d-3)/2$ with $j = 3/2, 5/2, 7/2,...$ as before, as well as
\begin{equation} \label{eq:RSRNterms}
A= \frac{d-2}{2} \sqrt{f(r)} + \left( \kappa + C \right) \hspace{0.2cm} \text{and} \hspace{0.2cm} B= \frac{d-2}{2} \sqrt{f(r)} - \left( \kappa + C \right) \;,
\end{equation}
\noindent it may be shown that
\begin{eqnarray}
W &=& \frac{d-3}{AB} \sqrt{\frac{f(r)}{r^2}} \left[ \frac{2AB}{d-2} \left(\kappa + C \right) + \frac{d-2}{2}\left((1-f(r))\kappa + C \right)\right] - \frac{d-4}{d-2} \sqrt{\frac{f(r)}{r^2}} \left(\kappa + C \right)  \nonumber \\
 & = & \frac{\sqrt{f(r)}}{r} \; \left(\kappa + C \right) \left[1 +\frac{(d-3)(d-2)}{2(\kappa + C)} \frac{(1-f(r))\kappa + C}{\frac{(d-2)^2}{4} f(r)-\left(\kappa + C \right)^2} \right] \label{eq:WnonTTRSRN} \;.
\end{eqnarray}
\noindent Though not obvious by inspection, eq. ({\ref{eq:bigL}}) once again emerges when the potentials are subjected to the $\ell \rightarrow \infty$ limit.

\par Finally, we discuss the Schwarzschild-Tangherlini (A)dS BH \cite{refRSSAdS19}, for which a metric function of $f(r) = 1 - 2 \mu/r^{\;d-3} - \lambda r^{\;2}$ is utilised. We maintain the definitions of $\kappa$ and $\zeta$ for the non-TT and TT modes, respectively. However, the spin-3/2 effective QNM potentials within SAdS BH spacetimes are distinguished by their more complicated expressions, from which the QNF $\omega$ cannot be extracted. Let us first consider the non-TT eigenmodes, which produce
\begin{equation} \label{eq:RSSdSnontt}
V_{1,2} =\mp \partial_{r_*} W + W^2 \;, \hspace{0.5cm}  W = \left[ {\cal D}^2 - {\cal B}^2 \right]^{1/2} f^{-1} {\cal F} \;.
\end{equation}
\noindent Here, $\partial_{r_*} = {\cal F} \partial_r \;,$  for which we introduce
\begin{equation}
{\cal B} = i \kappa \frac{\sqrt{f(r)}}{r} \; (z+1) \hspace{0.5cm} \text{and} \hspace{0.5cm} {\cal D} = -i \sqrt{\lambda f(r)} \;\frac{(d-2)}{2} \; \left( z+ \frac{d-4}{d-2} \right) \;,
\end{equation}
as well as
\begin{equation} \label{eq:z}
z = \frac{1- \left(1 + \frac{(d-3)(d-2)}{2} \frac{2 \mu}{r^{d-3}}\right)}{\kappa^2 - \frac{(d-2)^2}{4} \left(1 - \frac{2 \mu}{r^{d-3}} \right)} \;,
\end{equation}
to define
\begin{equation}
{\cal F} = f(r) \left[ 1 + \frac{f(r)}{2 \omega} \left( \frac{\partial}{\partial r} \frac{{\cal D}}{i {\cal B}} \right) \left(\frac{ {\cal B}^2}{{\cal B}^2 - {\cal D}^2} \right) \right]^{-1} \;.
\end{equation}

\par Similarly, for the TT eigenmodes in $d\geq 5$,
\begin{equation}
\mathbb{V}_{1,2} =\mp \partial_{r_*} \mathbb{W} + \mathbb{W}^2  \;, \hspace{0.5cm} \mathbb{W} = \left[ \mathbb{D}^2 - \mathbb{B}^2 \right]^{1/2} f^{-1} \mathbb{F} \;,
\end{equation}
where $\partial_{r_*} = \mathbb{F} \partial_r$, and the introduction of
\begin{equation}
\mathbb{B} = i \zeta \frac{\sqrt{f(r)}}{r}  \hspace{0.5cm} \text{and} \hspace{0.5cm} \mathbb{D} =  -i \sqrt{\lambda f(r)} \;\frac{(d-2)}{2}
\end{equation}
suffices to define
\begin{equation}
\mathbb{F} = f(r) \left[ 1 + \frac{f(r)}{2 \omega} \left( \frac{\partial}{\partial r} \frac{{\mathbb D}}{i {\mathbb B}} \right) \left(\frac{ {\mathbb B}^2}{{\mathbb B}^2 - {\mathbb D}^2} \right) \right]^{-1} \;.
\end{equation}

\par Despite the unusual inclusion of the QNF within the potential, the application of $\kappa \rightarrow \infty \;$ and $\zeta \rightarrow \infty \;$ yield the desired outcome featured in eq. ({\ref{eq:bigkappa}}) for the SAdS cases, such that the overall behaviour of these effective potentials within the eikonal limit for the spin-3/2 QNMs in the SdS BH spacetimes reflects eq. (\ref{eq:bigL}). However, the $V_{eff} \rightarrow r^2$ for $r \rightarrow \infty$ behaviour associated with AdS spacetimes must be taken into account, such that the overall behaviour of the spin-3/2 effective QNM potential reflects eq. ({\ref{eq:bigLAdS}}) for the Schwarzschild-Tangherlini AdS BH.

\par Thus, for higher-dimensional stationary and spherically-symmetric BHs, the effective potentials for spin-3/2 perturbing fields reduce to the same expressions in the large angular momentum limit as their integer spin counterparts.

%
%

\section{\label{sec:num}Numerical analysis of the large multipolar regime for QNFs in 4D spherically-symmetric BH spacetimes}
\par While numerical techniques have been exploited extensively in the study of integer-spin QNFs (with some attention directed to the effect of the large-$\ell$ limit in Refs. \cite{refGoebel1972,refFerrariMashhoon,refShuShen,refCardosoLyapunov}, among others), spin-$1/2$ and spin-$3/2$ QNF calculations have not been discussed as widely. Known examples of the application of numerical techniques specifically to QNFs of half-integer fields include the spin-1/2 QNF results of Ref. \cite{refDirac03} and Ref. \cite{refDirac07}, where the 3rd-order WKB approximation was used for the Schwarzschild case in four and higher dimensions, respectively. Further work in this spacetime was conducted in Ref. \cite{refPanJing} and \cite{refBlazSal} with the CFM and WKB; Refs. \cite{refChakrabarti} and  \cite{refGiammatteo} instead explored the RN and SAdS spacetimes, respectively. Less common are studies on the spin-3/2 QNFs. From the gauge-invariant formalism constructed by some of the authors, results have been obtained for spin-3/2 QNFs via Konoplya's 6th-order WKB and improved AIM in Schwarzschild-Tangherlini contexts \cite{refRSSchwarz16}, via the CFM for the RN BH spacetime \cite{refRSRN18}, and via the Horowitz-Hubeny approach for S(A)dS cases with $d\geq5$ \cite{refRSSAdS19}. A major component of this section is thus to add to the extant results for the QNFs of half-integer fields by applying Dolan and Ottewill's inverse multipolar expansion method \cite{refDolanOttewill2009} to the spin-3/2 fields, as well as spin-1/2 fields in the RN context, for the first time.

\par Specifically, we introduce this numerical component to validate the analytic results of section \ref{sec:potentials}, to explore the behaviour of QNFs in the eikonal limit for different BH spaceimes, as well as to engage with the recently constructed method of Dolan and Ottewill $-$ which was found to be as efficient as Konoplya's 6th-order WKB \cite{refDolanOttewill2009,refDecanini2011}. In their original paper \cite{refDolanOttewill2009}, Dolan and Ottewill provide explicit expressions for their multipolar expansions to $\mathcal{O} (L^{-6})$ for integer-spin perturbations in the Schwarzschild context, to $\mathcal{O} (L^{-1})$ and $\mathcal{O} (L^{-4})$ for the gravitational perturbations of a general and extremal RN BH, respectively, and to $\mathcal{O} (L^{-4})$ for the scalar perturbations of a SdS BH. Here, we extend their results to encompass fields of $s \in \{0,1/2,1,3/2,2\}$ within these three spacetimes, to orders of $\mathcal{O} (L^{-6})$ in almost every case, with the objective of exploring the interplay between $\theta^2$, $\lambda$, and $\ell$ in the eikonal limit. We verify our results using expressions from the literature, where available, as well as Konoplya's 6th-order WKB method \cite{Konoplya2003} and the PT approximation of Refs. \cite{PoschTellerMethod,refFerrariMashhoon}. Such comparisons with the low-lying QNFs are available in appendix \ref{app}. 

\subsection{\label{subsec:DO} The Dolan-Ottewill expansion method}

\par Since its introduction in Ref. \cite{refDolanOttewill2009}, the Dolan-Ottewill expansion method (hereafter, the DO method) has proven itself a powerful tool whose efficiency in certain contexts (such as the Regge pole determination of Ref. \cite{refDecanini2011}) surpasses even that of the WKB method. It has been applied to massive scalar perturbations of a Schwarzschild BH \cite{refDecanini2011}, massless scalar perturbations of a RN and Bardeen BH  \cite{refFernandoCorrea}, massless electromagnetic perturbations of a RN BH \cite{refLiLinYang}, and massless Dirac fields of a number of spherically-symmetric regular BHs (none of which we pursue here) in Ref. \cite{refLiMaLin2013}, all with reasonable success even at relatively low orders. As mentioned previously, the method is designed to calculate QNFs through the application of a novel ansatz to eq. ({\ref{eq:ode}}), in conjunction with a multipolar expansion in orders of $L= \ell + 1/2$. The ansatz in question is constructed from an analysis of the critical orbits of null geodesics and serves to characterise the method. Unlike the WKB and PT, whose development centres on the form of the potential, the DO method relies almost entirely on the nature of the spacetime context. The method is especially appealing due to the relative ease with which it can be carried to very high orders
in a manner that is globally valid in $r$. This feature endows it with its high level of accuracy 
\cite{refDolanOttewill2009}.

\par That the method can be extended to the calculation of QNM wavefunctions, as well as the quasinormal excitation factors that are invaluable in a wide variety of applications within BH perturbation theory and gravitational-wave analysis \cite{refBertiCardoso,refDolanOttewill2011,refDolanOttewill2013}, serve to elevate its appeal and further motivate its study. Moreover, the method is pragmatic: since QNMs most likely to be observed are associated with massless perturbations \cite{refKonoplyaZhidenkoReview} and the ``fundamental" $n=0$ mode that dominates QNM spectra, the original construction of the method was geared towards use in observation.

\par Here, we provide a brief overview of the method and the physical concepts from which it is constructed, as outlined in Ref. \cite{refDolanOttewill2009}. Let us begin with the metric for a static, spherically-symmetric, $d$-dimensional BH spacetime,
\begin{equation} \label{eq:metric}
ds^2 = -f(r)dt^2 + f(r)^{-1} dr^2 + r^2 d \Omega_{d-2} \;.
\end{equation}
\noindent To extract the parameters of interest, we follow Refs. \cite{Chandrasekhar1983,refCardosoLyapunov} in studying the equation of motion for a test particle near the spherically-symmetric BH. The Lagrangian in the equatorial plane ($\theta = \pi/2$) is written as
\begin{equation} \label{eq:Lagrangian}
{\cal L} = \frac{1}{2} g_{\mu \nu} \dot{x}^{\mu} \dot{x}^{\nu} = \frac{1}{2} \left( -f(r) \dot{t}^2 + f(r)^{-1} \dot{r}^2 + r^2 \dot{\phi}^2 \right) \;,
\end{equation}
\noindent where the overdot represents a derivative with respect to an affine parameter. From the corresponding conjugate momenta,
\begin{eqnarray}
p_t & = & f(r) \dot{t} \equiv E \label{eq:E} \;, \\
p_{\phi} & = & r^2 \dot{\phi} \equiv L \label{eq:L} \;,\\
p_r & = & f(r)^{-1} \dot{r} \;,
\end{eqnarray}
\noindent we obtain
\begin{equation}
\dot{\phi} = \frac{L}{r^2} \;, \hspace{1.5cm} \dot{t} = \frac{E}{f(r)} \;.
\end{equation}
\noindent These expressions allow us to write the Hamiltonian as
\begin{eqnarray}
{\cal H} & = & \left( p_t \dot{t} + p_{\phi} \dot{\phi} + p_r \dot{r} - {\cal L} \right) \nonumber \\
& \Rightarrow & 2 {\cal H} = E \dot{t} - L\dot{\phi} - f(r)^{-1} \dot{r}^2 = \delta_1 \;,
\end{eqnarray}
\noindent for which $\delta_1 = 1$ for time-like geodesics and $\delta_1 = 0$ for null geodesics. Our interest lies in the latter. The definition $\dot{r} \equiv V_r$ \cite{refCardosoLyapunov} then permits the expression,
\begin{equation}  \label{eq:vr}
V_r = f(r) \left[ \frac{E^2}{f(r)} - \frac{L^2}{r^2} - \delta_1 \right] \;.
\end{equation}
\noindent This, combined with the condition for circular orbits $V_r = V_r' = 0$ \cite{refBardeenPressTeukolsky}, yields
\begin{equation}
0 = \frac{E^2}{L^2} - \frac{f(r)}{r^2}
\end{equation}
\noindent for the null geodesics. The radius of this circular orbit can then be obtained by solving
\begin{equation} \label{eq:rc}
r_c = \frac{2 f_c}{\partial_r f_c} \;,
\end{equation}
\noindent where the subscript $c$ denotes evaluation at $r=r_c$ throughout this section, {\textit{viz.}} $f_c = f(r_c)$.

\par Dolan and Ottewill make use of the ``impact parameter" $b = L/E$ to express the above as a function
\begin{equation}
k^2(r,b) = \frac{1}{b^2} - \frac{f(r)}{r^2} \;.
\end{equation}
\noindent When evaluated at $ \{ r_c, b_c \}$, where $b=b_c$ serves as the ``critical impact parameter", $k^2 (r, b_c)$ has a repeated root  and meets the conditions,
\begin{equation}
k^2 (r_c, b_c) = \partial_r k^2 (r_c, b_c) = 0 \;.
\end{equation}
\noindent Dolan and Ottewill then make the explicit assumption that the repeated root is a double root, in order to produce
\begin{equation} \label{eq:k}
k_c (r) = \text{sgn} (r- r_c) \sqrt{k^2 (r, b_c)} = (r-r_c) K(r) \;.
\end{equation}
\noindent The inclusion of $\text{sgn} (r-r_c)$ allows for the specification that $k_c > 0$ for $r>r_c$. As discussed in section 5.2 of Ref. \cite{refFernandoCorrea}, the sign function is incorporated to ensure that $k_c(r)$ is differentiable at $r=r_c$. $K(r)$ can be used in the definition of the Lyapunov exponent \cite{refDolanOttewill2009,refCardosoLyapunov},
\begin{equation}
\overline{\Lambda} = \sqrt{ \frac{\partial^2_r V_{r_c} }{2 \dot{t}^2}} = f_c K_c \;,
\end{equation}
\noindent where $V_{r_c}$ refers to eq. (\ref{eq:vr}) evaluated at $r=r_c$.

\par With this set up in place, we can proceed to the procedure itself. First, we must restructure eq. (\ref{eq:ode}) in terms of $L = \ell + (d-3)/2$. For the case of the $d$-dimensional scalar field,
\begin{equation} \label{eq:odescalar}
\frac{d^2 \psi}{dr_*^2} + \left[ \omega^2 - \frac{f(r)}{r^2}\left( L^2 - \frac{(d-3)^2}{4} \right) - f(r)V_{eff} (r) \right] \psi = 0 \;,
\end{equation}
\noindent for which $V_{eff}$ is understood to depend on $L$ as
\begin{equation}
V_{eff} (r) = \sum_{k=-1}^{\infty} V(r)_k L^{-k}  \;.
\end{equation}

\par It is in this equation that we are to introduce the novel ansatz, defined in terms of the critical impact parameter and eq. (\ref{eq:k}),
\begin{equation} \label{eq:ansatz}
\psi(r)= \exp \bigg \{ \int^{r_{*}} b_{c}k_{c}(r) dr_{*} \bigg \} \; v(r)  \;,
\end{equation}
\noindent which we can write as $\psi(r)= \exp \{ i \omega z(r_*) \} v(r)$ using  $z(r_*) = \int^{r_{*}} \rho(r) dr_{*}$ for $\rho(r)=b_{c}k_{c}(r)$.
\noindent To encapsulate the ingoing and outgoing boundary conditions required, it is assumed that
\begin{eqnarray}
f(r) \rightarrow 0, \; \; \; b_{c}k_{c}(r) \rightarrow -1 & \text{as} & \; r_* \rightarrow - \infty \;, \\
f(r)/r^2 \rightarrow 0, \; \; b_{c}k_{c}(r) \rightarrow +1 & \; \; \; \;  \text{as} \; \; \; \; & \; r_* \rightarrow + \infty  \;.
\end{eqnarray}
\noindent This relays that a horizon is encountered as $r_* \rightarrow - \infty$, while $r_* \rightarrow + \infty$ leads either to an asymptotically flat region or a cosmological constant. The boundary conditions upon which the DO method depends, therefore, accommodates asymptotically flat and dS spacetime, but not AdS contexts.

\par Once these parameters are defined, the ansatz is substituted into the recasted ordinary differential equation, and the expression
\begin{equation} \label{eq:ode4}
 f(r) \frac{d}{dr} \left( f(r) \frac{dv}{dr} \right) + 2i \omega \rho(r)\frac{dv}{dr} +  \left[  i \omega  f(r) \frac{d \rho}{dr} +  ( 1 - \rho(r)^2) \;\omega^2  - V_{eff}(r) \right]v(r) = 0
\end{equation}
\noindent may be obtained (generalising eq. (6) of Ref. \cite{refDolanOttewill2009} and offering a possible correction to the second term of eq. (40) of Ref. \cite{refFernandoCorrea}, eq. (21) in Ref. \cite{refLiLinYang}, and eq. (31) in Ref. \cite{refLiMaLin2013} that may allow for computation to higher orders of $L^{-k}$). The objective is then simply to solve for the QNFs at each order of $L^{-k}$ through recursive substitutions of
\begin{equation} \label{eq:omegaseries}
\omega= \sum_{k=-1}^{\infty} \omega_k L^{-k}
\end{equation}
\noindent and
\begin{equation} \label{eq:wavefunction}
v(r)= \exp \bigg \{ \sum_{k=0}^{\infty}  S_k(r) L^{-k} \bigg \} \;,
\end{equation}
\noindent where these expansions hold for the least damped ``fundamental mode" of $n=0$ \cite{refDolanOttewill2009} which dominates the QNF spectrum,

\par Specifically, we solve iteratively for $\omega_k$ and $S'_k (r)$, the results of which are substituted into eq. (\ref{eq:omegaseries}) to determine the QNF. The associated wavefunction may be obtained by integrating the $S'_k (r)$ terms and substituting these into eq. ({\ref{eq:wavefunction}}). We reserve an investigation into the wave functions for a future work, and focus only on the QNFs in this section.
\par The DO method is thus heavily reliant on the nature of the BH spacetime and its associated photon orbits: the radius of the critical orbit $r_c$, the impact parameter $b_c$, and the introduced $k_c(r)$ function serve as the cruces of the approach. This physicality emerges even from the terms of the QNFs: the first QNF term for which we solve (occurring at order $L^{+2}$) can be defined using $1-b_c^2 k_c^2(r) =  b_c^2 f(r) / r^2$ and produces
\begin{equation} \label{eq:wneg1}
\omega_{-1} =  \frac{1}{b_c} = \Omega_c \;.
\end{equation}
\noindent Here, $\Omega_c$ is the orbital angular velocity evaluated at $r=r_c$. Moreover, the expansion in inverse powers of $L$ affixes angular dependence on the QNF, which has only an implicit dependence on $\ell$. Recall from eq. (\ref{eq:Lyapunov}) that $\Omega_c \approx \overline{\Lambda}$ in the large multipolar limit.

\par We now proceed to a complete study of the DO method for QNFs of spin $s \in \{ 0,1/2,1,3/2,2\}$ in the 4D Schwarzschild, RN, and SdS BH spacetimes. We apply the method also to the 	``uniform" potential of eq. (\ref{eq:bigL}): since this expression encodes the generalised behaviour of fields in the large multipolar limit, it is useful to observe a direct application of numerical methods to this expression when analysing the behaviour of individual fields in the large-$\ell$ limit. The calculations themselves were performed within the Mathematica environment (versions 8 and 11.2) through an algorithm we produced based on Ref. \cite{refDolanOttewill2009}. We did not explicitly set a precision assignment or compute the associated errors for the input parameters within our numerical process; instead, we made use of Mathematica's internal error setting. In our implementation, we first specify eq. (\ref{eq:ode4}) for the spacetime of interest, with the substitution of eqs. (\ref{eq:omegaseries}) and (\ref{eq:wavefunction}) expanded to the desired order. The first value of eq. (\ref{eq:omegaseries}), $\omega_{-1}$, must be solved for and substituted into eq. (\ref{eq:ode4}), followed by $\omega_{0}$ and $S'_0(r)$. Through iterative solving and substituting of $\omega_k$ and $S'_k(r)$ into eq. (\ref{eq:ode4}) for increasing values of $k$, we are able to produce an expression for the QNF as an expansion in inverse multipolar numbers. To specify the spin of the field, the appropriate expression for the effective potential must be substituted into eq. (\ref{eq:ode4}) from the onset. It is worth noting that for RN BH spacetimes, the value of $\theta$ must also be specified.

\par We note also that our treatment of QNFs associated with fields of integer and half-integer spin differ slightly: when applying the DO method to integer fields, we parametrise the angular momentum as $L = \ell + 1/2$; for the half-integer fields, we use $\bar{L}$ instead. This change in parametrisation is necessitated by the differing representation of the angular momentum within the effective potentials for integer and half-integer fields, as shown in section \ref{sec:potentials}. Further distinction is also required for the definition of $\bar{L}$ for spin-1/2 and spin-3/2 fields. For the Dirac case (and the ``uniform" potential),
\begin{center}
 $ \bar{L} = L + 1/2 = \kappa = j + 1/2 \equiv \ell + 1 $ for $\ell \in \mathbb{N}_o$ and $j = 1/2,3/2,... $ \cite{refDirac07,refDirac03,refZhidenko2004}.
\end{center}
\noindent For the Rarita-Schwinger case, 
\begin{center}
 $ \bar{L} = L+1/2 = \kappa = j + 1/2 \equiv \ell + 2$ for $\ell \in \mathbb{N}_o$ and $j = 3/2,5/2,...$ \cite{refRS15,refRSSchwarz16,refRSRN18,refRSSAdS19}.
\end{center}
\noindent These are in accordance with the definitions of the spinor eigenvalue on the $(d-2)$-sphere provided in section \ref{sec:potentials}. For further clarity on the choice of parametrisations, see appendix \ref{app}. Consider Tables \ref{table:SdSGravLow} and \ref{table:SdSDiracLow} for the distinction between $L$ and $\bar{L}$, where the results for spin-2 QNFs can be compared with those of spin-1/2 QNFs, respectively. For the distinction between half-integer cases, consider the spin-1/2 QNFs of Table \ref{table:SchwarzComp1} and the spin-3/2 QNFs of Table \ref{table:SchwarzComp2}. 

\par We maintain these parametrisations when computing the QNFs throughout this work. Furthermore, we observe that when calculating the QNFs within Mathematica, it is more convenient to divide eq. (\ref{eq:ode4}) and the effective potentials by $f(r)$. To verify our large-$\ell$ results, we use the 6th-order WKB method of Ref. \cite{Konoplya2003} and the PT approximation of Refs. \cite{PoschTellerMethod,refFerrariMashhoon} (guided by the publicly available Mathematica notebooks from Refs. \cite{Konoplya2003} and \cite{refZhidenko2004}). The low-lying QNFs included in appendix \ref{app} serve as further sources of validation, where extant results in the literature computed via various established numerical and analytical techniques compare favourably with our results produced via the DO method. 

\par Finally, since we confine ourselves to the 4D context, we do not consider TT eigenmodes for the Rarita-Schwinger field, tensor modes of the gravitational field, nor scalar modes of the electromagnetic field. Where possible, we can also exploit the isospectrality of the QNFs to describe their behaviour in the large-$\ell$ limit, {\textit{viz.}} the scalar and vector modes of the gravitational perturbation \cite{refIsospectralRWZ} and the two parities of the half-integer fields \cite{refSUSYpot}. We set the BH mass as $\mu = 1$ throughout the work.

\subsection{\label{subsec:schwarz} QNF behaviour within the large-$\ell$ limit for Schwarzschild BHs}

\par For the metric function $f(r)=1-2/r$,
\begin{equation}
r_c =  3 \;, \hspace{0.4cm} b_c = \sqrt{27} \; \hspace{0.4cm} \Rightarrow \; \; \rho (r) = \left(1-\frac{3}{r} \right) \sqrt{1+\frac{6}{r}} \;\;. \label{eq:SchwarzSpecs}
\end{equation}
\noindent With these components inserted into eq. ({\ref{eq:ode4}}), the DO method can be used to solve for the appropriate QNF. These expressions are enclosed in Table \ref{table:DOschwarz} up to $\mathcal{O}(L^{-6})$. For the fields of integer spin, the effective potential used is that of eq. (\ref{eq:RWconcise}), with $\ell (\ell + 1) \rightarrow L^2 - 1/4$. For the Dirac case, we used the positive parity and parametrised the effective potential as 
\begin{equation} \label{eq:DODiracSchwarz}
V(r) =  f(r) \frac{d}{dr} \left( \frac{\sqrt{f(r)}}{r} \right) {\bar{L}} + \frac{f(r)}{r^2} {\bar{L}}^2\;.
\end{equation}
\noindent For the Rarita-Schwinger effective potential, the superpotential of eq. (\ref{eq:WnonTTRSschwarz}) reduces to
\begin{equation}
W \bigg \vert_{d=4} = \frac{\sqrt{f(r)}}{r} \kappa  \left[ \frac{\kappa^2-1}{\kappa^2 - f(r)} \right] = \frac{\sqrt{f(r)}}{r} \kappa  \left[1 + \frac{1-f(r)}{f(r) - \kappa^2} \right]
\end{equation}
\noindent in 4D. With the $\kappa \rightarrow \bar{L}$ parametrisation in place, we observe that eq. (\ref{eq:RSScwarznontt}) becomes
\begin{equation} \label{eq:DORSSchwarz}
V(r) = f(r) \frac{d}{dr} \left( \frac{\sqrt{f(r)}}{r} \; [1 + z_s] \right) {\bar{L}} + \frac{f(r)}{r^2} \;[1+z_s]^2 {\bar{L}}^{\;2} \;,
\end{equation}
\noindent in 4D, where
\begin{equation} \label{eq:zs}
z_s = \frac{1-f(r)}{f(r) - \bar{L}^2} \;.
\end{equation}
\noindent Since the TT modes emerge only for $d \geq 5$, the spin-3/2 behaviour is wholly captured by this non-TT mode in 4D. To compute the Rarita-Schwinger QNFs, we expand $z_s$ in inverse powers of $\bar{L}$,
\begin{equation}
z_s \approx - \frac{2}{r \bar{L}^2} - \frac{2(r-2)}{r^2 \bar{L}^4} - \frac{2 (r-2)^2}{r^3 \bar{L}^6} \;.
\end{equation}
\noindent Note that if we set $z_s\rightarrow 0$, we recover the Dirac potential. 

\par We run the generalised DO programme we construct from eqs. (\ref{eq:ode4}) and ({\ref{eq:SchwarzSpecs}}) within Mathematica, for each potential defined above, to generate the QNF in the form of eq. ({\ref{eq:omegaseries}}). These inverse multipolar expansions are recorded in Table \ref{table:DOschwarz}. Within, certain commonalities emerge among different $s$, such as a negative $L^0$ term and consistently imaginary expressions for $L^{-k}$ terms of even $k$. To express the QNFs provided in Tables \ref{tableL1} and \ref{tableL2}, we extract $\sum^6_{k=-1}  \omega_k L^{-k}$ only from Table \ref{table:DOschwarz} and substitute the appropriate values of $L$ and $\bar{L}$, respectively. In our comparison of our results with those we calculate via the WKB and PT methods, we observe excellent agreement and an almost perfect correlation with the 6th-order WKB results to four decimal places from $\ell \thicksim 10$ onwards.

\begin{table}[t]
\centering
\caption{\textit{The inverse multipolar expansions for the effective QNFs of spin $s$, where integer-spin results are reproduced. All other expressions are new and derived using $L \rightarrow \bar{L}$. Here, the ``uniform" potential refers to eq. (\ref{eq:bigL}).}}
\begin{tabular}{| c | c |}
\hline
$s$ & $ b_c \sum^6_{k=-1}  \omega_k L^{-k} $ \\
\hline
 & \textit{perturbations of integer spin} \\
\hline
$0$ & $L -\frac{i}{2} + \frac{7}{216L} - \frac{137}{7776 L^2}i + \frac{2615}{1259712L^3} + \frac{590983}{362797056L^4}i -\frac{42573661}{39182082048L^5} + \frac{11084613257}{8463329722368 L^6} i $
 \\
$1$ & $L -\frac{i}{2} - \frac{65}{216L} +  \frac{295}{7776L^2}i - \frac{35617}{1259712L^3} + \frac{3374791}{362797056 L^4}i - \frac{342889693}{39182082048 L^5} + \frac{74076561065}{8463329722368 L^6} i $
 \\
$2$ & $L -\frac{i}{2} -\frac{281}{216L} + \frac{1591}{7776L^2}i -\frac{710185}{1259712L^3} + \frac{92347783}{362797056 L^4}i -\frac{7827932509}{39182082048 L^5} - \frac{481407154423}{8463329722368 L^6}i$\\
\hline
& \textit{the uniform potential}  \\
\hline
& $ {\bar{L}}-\frac{i}{2} -\frac{19}{108 {\bar{L}}} +\frac{295 }{7776 {\bar{L}}^2}i +\frac{3853}{2519424 {\bar{L}}^3} -\frac{66089 }{362797056 {\bar{L}}^4}i -\frac{165538573}{39182082048 {\bar{L}}^5} +
\frac{54780211001}{8463329722368 {\bar{L}}^6}i$\\
\hline
& \textit{perturbations of half-integer spin} \\
\hline
$1/2$ & ${\bar{L}} -\frac{i}{2} -\frac{11}{216 {\bar{L}}} -\frac{29 }{7776 {\bar{L}}^2}i +\frac{1805}{1259712 {\bar{L}}^3} +\frac{27223 }{362797056 {\bar{L}}^4}i +\frac{23015171}{39182082048 {\bar{L}}^5} -\frac{6431354863 }{8463329722368 {\bar{L}}^6}i $\\
 $3/2$ & ${\bar{L}} -\frac{i}{2} -\frac{155}{216 {\bar{L}}} +\frac{835 }{7776 {\bar{L}}^2}i -\frac{214627}{1259712 {\bar{L}}^3} +\frac{25750231 }{362797056 {\bar{L}}^4}i -\frac{2525971453}{39182082048 {\bar{L}}^5} + \frac{292606736465 }{8463329722368 {\bar{L}}^6}i  $ \\
\hline
\end{tabular}
\label{table:DOschwarz}
\end{table}

\par The interplay between QNM parameters for fields of different spins within the Schwarzschild BH spacetime was noted explicitly in Ref. \cite{refShuShen} through a study of $\ell \in [2,5]$: for fields of spin 0, 1, and 2, $\mathbb{R}e \{ \omega \}$ decreased with $s$ for fixed $n$ and $\ell$, but increased with the $s$ of half-integer fields; $\mathbb{I}m \{ \omega \}$ was found to decrease with $s$ for all fields. This behaviour reported for the real part of the QNFs is replicated in our results, and holds true as we increase $\ell$. The imaginary part in the large multipolar limit tends to a constant value, irrespective of the spin of the perturbing field.

\par That the imaginary part is negative indicates that the modes decay with time \cite{refBlazSal}. The magnitude of the imaginary part, however, remains consistently $\sim 0.0962i$ for fields of integer and half-integer spin within the large-$\ell$ limit. From this, we observe that the imaginary component of the QNF corresponds to eq. (\ref{eq:Lyapunov}) and matches the correct value for the Lyapunov exponent associated with the Schwarzschild BH irrespective of the spin of the perturbing field.

\begin{table}[t]
\centering
\caption{\textit{Scalar, electromagnetic, and vector-type gravitational QNFs for a 4D Schwarzschild BH calculated via the DO method to order $\mathcal{O}(L^{-6})$.}}
\begin{tabular}{| C | C | C | C |}
\hline
 \ell  &  \omega_{s} & \omega_{em} &  \omega_{g} \\
\hline
10  & 2.0213 - 0.0963i & 2.0152 - 0.0962i  & 1.9968-0.0959i \\
20  & 3.9455-0.0962i  & 3.9424- 0.0962i & 3.9330-0.0961i \\
40  & 7.7944- 0.0962i  & 7.7928-0.0962i  & 7.7880-0.0962i \\
60  & 11.6433 - 0.0962i & 11.6423 - 0.0962i & 11.6391-0.0962i \\
80  & 15.4923 -0.0962i & 15.4915 - 0.0962i & 15.4891-0.0962i \\
100 & 19.3413 -0.0962i  & 19.3407 - 0.0962i                                            & 19.3387-0.0962i \\
\hline
\end{tabular}
\label{tableL1}
\end{table}

\begin{table}[t]
\centering
\caption{\textit{Uniform (eq. (\ref{eq:bigL})), Dirac, and Rarita-Schwinger QNFs for a 4D Schwarzschild BH calculated via the DO method to order $\mathcal{O}(L^{-6})$.}}
\begin{tabular}{| C | C | C | C |}
\hline
\ell   &  \omega_{uni} & \omega_{D}   & \omega_{RS} \\
\hline
 10 & 2.1139-0.0962 i & 2.1161-0.0962 i & 2.2979-0.0961 i\\
 20 & 4.0398-0.0962 i & 4.0410-0.0962 i & 4.2276-0.0962 i\\
 40 & 7.8896-0.0962 i & 7.8902-0.0962 i & 8.0796-0.0962 i \\
 60 & 11.7389-0.0962 i & 11.7393-0.0962 i & 11.9297-0.0962 i\\
 80 & 15.5880-0.0962 i & 15.5883-0.0962 i & 15.7792-0.0962 i\\
 100 & 19.4371-0.0962 i & 19.4374-0.0962 i & 19.6286-0.0962 i\\
\hline
\end{tabular}
\label{tableL2}
\end{table}

\par Though Pan and Jing reported an equidistant spacing of
\[\mathbb{ R}e \{ \Delta \omega \} \approx 0.3849  \;, \]
\noindent between successive values of $\omega$ for large $\ell$ in Ref. \cite{refPanJing}, we find instead that
\begin{equation}
\mathbb{ R}e \{ \Delta \omega \}  \approx 0.1925 \pm 0.0001 \;.
\end{equation}
\noindent Recall from eq. ({\ref{eq:Lyapunov}) that this value is equivalent to the Lyapunov exponent for the Schwarzschild BH spacetime. In the large multipolar limit, however, $\overline{\Lambda} \approx \Omega_c$, such that this spacing is related also to the orbital angular frequency. This value of $\mathbb{ R}e \{ \Delta \omega \}$ first arises at $\ell \thicksim 10$; the increase in $\omega$ continues indefinitely as this spacing remains constant. The uniform potential responsible for providing a generalisation of the QNF behaviour in the large-$\ell$ limit acts as an average of the fields, and carries the same observed trends and relationship with the Lyapunov exponent.

\par Finally, we note that the integer-spin QNFs converge to a value of $\mathbb{R}e \{ \omega \} \approx 19.34$ for $\ell=100$; the Dirac and uniform QNFs tend to $\mathbb{R}e \{ \omega \} \approx 19.344$ while the Rarita-Schwinger QNFs become $\mathbb{R}e \{ \omega \} \approx 19.63$. This increasing uniformity between QNFs as $\ell$ is increased implies that the spin of the field is suppressed in the wake of high angular momentum, in a manner not unusual in astrophysical systems.

\subsection{\label{subsec:RN} QNF behaviour within the large-$\ell$ limit for RN BHs}
\par For the metric function $f(r) = 1 - 2/r + \theta^2/r^2$,
\begin{equation} \label{eq:RNspecs}
r_c = \frac{3 \pm \alpha}{2} \;, \hspace{0.5cm}b_c = \sqrt{\frac{(\alpha+3)^3}{2(\alpha+1)}} \hspace{0.4cm} \Rightarrow \; \; \rho (r) = \left(1-\frac{r_c}{r} \right) \sqrt{1+\frac{(\alpha -3)}{(\alpha + 1)} \left( \frac{r_c}{r} \right)^2 + \frac{(\alpha + 3)}{r} } \;.
\end{equation}
\noindent for $\alpha = \sqrt{9-8\theta^2}$ and using the outer orbit.

\par As in the Schwarzschild case, we use eq. ({\ref{eq:ZMconcise}}) to calculate the QNFs for the integer-spin perturbations with the $\ell (\ell + 1) \rightarrow L^2 - 1/4$ replacement. This carries over to the square root term included in the electromagnetic and gravitational perturbations,  such that $\ell(\ell+1) -2 \rightarrow L^2-9/4 \;.$
\noindent To apply the DO expansion successfully, however, an approximation becomes necessary. Though a first-order approximation of $q_{1,2} \approx 3 \mp 2L$ is suggested in Ref. \cite{refDolanOttewill2009}, we refine it to
\begin{equation} \label{eq:qRN}
q_{1,2} = 3 \mp \sqrt{9 + 4 \theta^2 \left( L^2 - \frac{9}{4} \right) \;} \; \approx \; 3 \mp 2 L \sqrt{\theta^{2}  +  y}, \hspace{0.3cm} y= \frac{9}{4L^2} (1 - \theta^2)
\end{equation}
\noindent in order to approach the higher-order multipolar expansions. For a small $y$, this becomes an expansion in inverse powers of $L$, and provides the condition
\[ L^{2}>\frac{9(1-\theta^{2})}{4\theta^{2}} \;,\]
\noindent which tells us that when $\theta$ is small, $L$ is more influential.
\par Note that this condition is not automatically suitable for $\theta=0$, as this forces $L$ to infinity. However, if we substitute $y$ into the square root first, then the Schwarzschild expression is recovered. Furthermore, we obtain the suggested $q_{1,2} \approx 3 \mp 2L$ when $\theta=1$. The DO method therefore does not fail for $\theta = 1$ $-$ in fact, it performs especially well then $-$ but instead breaks down for $\alpha = 0$ (i.e. for $\theta = \sqrt{9/8}\;$) \cite{refDolanOttewill2009}.

\par For the half-integer fields, eq. (\ref{eq:DODiracSchwarz}) with the inclusion of the RN metric function holds for the application of the DO method in the Dirac case. For the spin-3/2 field, we maintain the use of eq. (\ref{eq:WnonTTRSRN}), such that we write the superpotential as
\begin{equation}
W = \frac{\sqrt{f(r)}}{r} (1 + z_{_{RN}}) \;(\bar{L} + C) \;,
\end{equation}
\noindent where the charge term $C$ is defined in eq. ({\ref{eq:RSRNtermsC}}) and eq. ({\ref{eq:zs}}) is amended to
\begin{equation} \label{eq:zrn}
z_{_{RN}} = \frac{1}{\left( \bar{L} + C \right)} \; \frac{\left(1 - f(r) \right)\bar{L} + C }{f(r)- (\bar{L} + C )^2 } \;.
\end{equation}
\noindent As in the Schwarzschild case, $z_{_{RN}}$ is expanded in inverse powers of $\bar{L}$.

\par To investigate the effect of $\theta$ on the QNFs within the eikonal limit, we apply the DO method to perturbations of $s \in \{0,1/2,1,3/2,2 \}$, expanding to orders of $\mathcal{O}(L^{-6})$ unless otherwise stated. For each field, we first gauge the accuracy of the method in the RN BH context through a comparison with extant results (see appendix \ref{app:RN}) before proceeding with the large-$\ell$ investigation.

\par Please note that for the sake of brevity, we do not include the explicit $b_c \sum_k \omega_k L^{-k}$ expansions as in Table \ref{table:DOschwarz}: in our DO routine for the RN BH spacetime, we found that QNF computation was only feasible if $\theta$ was predefined. As such, a multipolar expansion for each $s$ is associated with each $\theta$; to calculate the QNF associated with a particular charge, the entire DO routine must be run. This undermines the efficiency of the method for charged BH spacetimes.

\par However, since the DO method breaks down when $\alpha = 0$ (i.e. for $\theta = \sqrt{9/8}$ rather than for $\theta=1$), it can be used where other numerical techniques fail, such as in the calculation of QNFs for the case of the extremal RN BH in asymptotically flat spacetimes where $\mu = \theta$. We show this explicitly by providing results for $\mu = \theta = 1$ in Tables \ref{table:RNIntegerLarge} and \ref{table:RNHalfIntegerLarge}, as well as in the comparisons made with the literature contained in appendix \ref{app:RN}. 

\par For the analysis of the QNFs in the RN BH spacetime, we consider the integer and half-integer perturbations separately. Observations made in the Schwarzschild case, however, manifest here: $\mathbb{R}e \{\omega \}$ increases indefinitely while $\mathbb{I}m \{ \omega \}$ tends to a constant related to the Lyapunov exponent $-$ in this case, a different constant for each $\theta$ $-$ and for fixed $\ell$, the magnitude of the QNF decreases (increases) for QNFs of increasing integer (half-integer) spin. Once again, an equidistant spacing appears between successive values of $\mathbb{R}e \{ \omega \}$ with respect to $\ell$, associated here with each $\theta$ and recorded in Table \ref{table:RNspacing}. We note also that when $\theta = 0$, the Schwarzschild QNFs are recovered. When $\theta \neq 0$, however, a discrepancy arises: the magnitudes of the real part of the QNFs of different spins do not converge to an approximately equal value, as we had seen in the Schwarzschild case.

\par Finally, we observe that our results remain in very good agreement with the QNFs calculated via the WKB method of Ref. \cite{Konoplya2003} and the PT approximation of Refs. \cite{PoschTellerMethod,refFerrariMashhoon} for $\theta <1$. 

\subsubsection{\label{subsec:RNinteger} QNFs of integer spin within 4D RN BH spacetimes}
\par In Table \ref{table:RNIntegerLarge}, we observe that for a fixed $\ell$, the real part of the QNF increases with $\theta$. This effect becomes pronounced for the electromagnetic QNFs: their increase with $\theta$ is the most rapid of the integer spin fields (a growth of $\sim 0.1$ more than the other QNFs for lower $\ell$). The imaginary part decreases with increasing $\theta$. 

\par For this electromagnetic case, we calculate the expansions up to $\mathcal{O}(L^{-4})$ for $\theta <1$. Beyond this order, our DO routine fails to produce reliable output. For $\theta =1$, however, we found that the DO method may be expanded to $\mathcal{O}(L^{-6})$ and further with ease. This reinforces the usefulness of the DO method: based on the results of Refs. \cite{refBHWKB3} and \cite{refGunter1980RN}, as well as our own WKB and PT calculations, standard methods may not necessarily accommodate $\theta  \rightarrow 1$ for $s=1$ QNF calculations. The DO method, on the other hand, proves very accurate for $\theta = 1$ and thereby proves itself a useful contribution to the QNF computational toolkit.

\par As explained in Ref. \cite{refBHWKB3}, the gravitational QNFs of the RN BH are inherently different from those of the Schwarzschild case: as a consequence of the coupling between electromagnetic and gravitational perturbations, only when $\theta = 0$ are the QNFs distinctly electromagnetic (such that $q_1$ applies) or gravitational (such that $q_2$ applies). That electromagnetic and gravitational perturbations exhibit a closely-related behaviour can be gleaned from their mathematical treatment in eqs. (\ref{eq:ZMconcise}) and (\ref{eq:qRN}). Consequently, the comments made for the electromagnetic
perturbations largely apply here: we calculate the expansions up to $\mathcal{O}(L^{-4})$ and observe once again that beyond this order, the DO method appears to break down for non-integer $\theta$. For the extremal case of $\theta = 1$, we produce QNFs to order $\mathcal{O}(L^{-6})$. Unlike in the spin-1 case, however, the gravitational QNF does not experience that same rapidity in growth for fixed $\ell$ and increasing $\theta$ $-$ we surmise that this behaviour is then unique to the electromagnetic QNF.
\vfill

\begingroup
\begin{table}[t]
\caption{\textit{QNFs of integer spin for a 4D RN BH calculated via the DO method to order $\mathcal{O}(L^{-4})$ for $\theta<1$ in the case of spin-1 and spin-2 fields, and to order $\mathcal{O}(L^{-6})$ otherwise.}}
\begin{tabular}{|C|C|C|C|C|C|}
\hline
field & \ell & \omega \; (\theta = 0.0) & \omega \; (\theta = 0.4) &  \omega \; (\theta = 0.8) & \omega \; (\theta = 1.0) \\
\hline
		&10 & 2.0213-0.0963 i & 2.0793-0.0971 i & 2.4385-0.0963 i & 2.6372-0.0882 i \\
		& 20 & 3.9455-0.0962 i & 4.0577-0.0970 i & 4.7501-0.0963 i & 5.1313-0.0883 i \\
scalar & 40 & 7.7944-0.0962 i & 8.0154-0.0970 i & 9.3786-0.0963 i & 10.1282-0.0884 i \\
		& 60 & 11.6433-0.0962 i & 11.9734-0.0970 i & 14.0084-0.0963 i & 15.1270-0.0880 i \\
		& 80 & 15.4920-0.0962 i & 15.9310-0.0970 i & 18.6390-0.0963 i & 20.1270-0.0884 i \\
		& 100 & 19.3410-0.0962 i & 19.8890-0.0970 i & 23.2690-0.0963 i & 25.1260-0.0884 i \\
\hline
		& 10 & 2.0152-0.0962 i & 2.0922-0.0973 i & 2.3667-0.0986 i & 2.7376-0.0882 i \\
		& 20 & 3.9424-0.0962 i & 4.0773-0.0972 i & 4.5730-0.0983 i & 5.2435-0.0883 i \\
 
EM		& 40 & 7.7928-0.0962 i & 8.0388-0.0971 i & 8.9761-0.0981 i & 10.2467-0.0884 i \\
		& 60 & 11.6423-0.0962 i & 11.9980-0.0971 i & 13.3768-0.0981 i & 15.2480-0.0884 i \\
		& 80 & 15.4920-0.0962 i & 15.9570-0.0971 i & 17.7770-0.0980 i & 20.2480-0.0884 i \\
		& 100 & 19.3410-0.0962 i & 19.9150-0.0971 i & 22.1770-0.0980 i & 25.2490-0.0884 i \\
\hline
		& 10 & 1.9968-0.0959 i & 2.0342-0.0964 i & 2.2236-0.0969 i & 2.4863-0.0882 i \\
		& 20 & 3.9330-0.0961 i & 4.0217-0.0968 i & 4.4310-0.0974 i & 4.9932-0.0883 i \\
	grav.	& 40 & 7.7880-0.0962 i & 7.9837-0.0969 i & 8.8343-0.0977 i & 9.9966-0.0884 i \\
		& 60 & 11.6391-0.0962 i & 11.9431-0.0970 i & 13.2351-0.0978 i & 14.9980-0.0884 i \\
		& 80 & 15.4890-0.0962 i & 15.9020-0.0970 i & 17.6350-0.0978 i & 19.9980-0.0884 i \\
		& 100 & 19.3390-0.0962 i & 19.8600-0.0970 i & 22.0350-0.0978 i & 24.9990-0.0884 i \\
\hline
\end{tabular}
\label{table:RNIntegerLarge}
\end{table}
\endgroup

\subsubsection{\label{subsec:RNHalfInteger} QNFs of half-integer spin within 4D RN BH spacetimes}
\par As in the integer-spin cases, the real (imaginary) part of the QNFs increases (decreases) with increasing $\theta$. However, the spin-3/2 QNFs are most affected by $\theta$, with a growth more rapid than even that of the spin-1 QNFs for lower $\ell$ ($\sim 0.2$ larger than the Dirac QNFs).

\begingroup
\begin{table}[t]
\caption{\textit{QNFs of half-integer spin and the ``uniform" potential for a 4D RN BH calculated via the DO method to order $\mathcal{O}(L^{-6})$ for all perturbations.}}
\begin{tabular}{|C|C|C|C|C|C|}
\hline
field & \ell & \omega \; (\theta = 0.0) & \omega \; (\theta = 0.4) &  \omega \; (\theta = 0.8) & \omega \; (\theta = 1.0) \\
\hline
 &10 & 2.1139-0.0961 i & 2.1739-0.0969 i & 2.4208-0.0979 i & 2.7475-0.0885 i \\
 & 20 & 4.0398-0.0962 i & 4.1544-0.0970 i & 4.6200-0.0979 i & 5.2487-0.0884 i \\
uniform & 40 & 7.8896-0.0962 i & 8.1132-0.0970 i & 9.0192-0.0979 i & 10.2493-0.0884 i \\
 & 60 & 11.7389-0.0962 i & 12.0715-0.0970 i & 13.4186-0.0979 i & 15.2500-0.0884 i \\
 & 80 & 15.5880-0.0962 i & 16.0300-0.0970 i & 17.8180-0.0979 i & 20.2500-0.0884 i \\
 & 100 & 19.4370-0.0962 i & 19.9889-0.0970 i & 22.2170-0.0979 i & 25.2500-0.0884 i \\
\hline
		& 10 & 2.1161-0.0962 i & 2.1761-0.0970 i & 2.4189-0.0979 i & 2.7489-0.0884 i \\
		& 20 & 4.0410-0.0962 i & 4.1555-0.0970 i & 4.6190-0.0979 i & 5.2494-0.0884 i \\
Dirac		& 40 & 7.8902-0.0962 i & 8.1138-0.0970 i & 9.0187-0.0979 i & 10.2497-0.0884 i \\
		& 60 & 11.7393-0.0962 i & 12.0719-0.0970 i & 13.4183-0.0979 i & 15.2500-0.0884 i \\
		& 80 & 15.5880-0.0962 i & 16.0300-0.0970 i & 17.8180-0.0979 i & 20.2500-0.0884 i \\
		& 100 & 19.4370-0.0962 i & 19.9880-0.0970 i & 22.2170-0.0979 i & 25.2500-0.0884 i \\
\hline
		& 10 & 2.2979-0.0961 i & 2.4022-0.0970 i & 2.6974-0.0985 i & 3.1099-0.0883 i \\
		& 20 & 4.2276-0.0962 i & 4.3813-0.0970 i & 4.9031-0.0983 i & 5.6167-0.0884 i \\
RS		& 40 & 8.0796-0.0962 i & 8.3393-0.0970 i & 9.3060-0.0981 i & 10.6206-0.0884 i \\
		& 60 & 11.9297-0.0962 i & 12.2974-0.0970 i & 13.7067-0.0981 i & 15.6220-0.0884 i \\
		& 80 & 15.7790-0.0962 i & 16.2550-0.0970 i & 18.1070-0.0980 i & 20.6230-0.0884 i \\
		& 100 & 19.6290-0.0962 i & 20.2140-0.0970 i & 22.5070-0.0980 i & 25.623-0.0884 i \\
\hline
\end{tabular}
\label{table:RNHalfIntegerLarge}
\end{table}
\endgroup

\par The computation of QNFs of half-integer spin is comparatively uncomplicated using the DO method, as there is no need to include further approximations than those specified for the Dirac and Rarita-Schwinger cases. For the uniform potential defined in eq. (\ref{eq:bigL}), we find that its behaviour aligns closest with that of the Dirac field.

\par Irrespective of the spin of the field, two distinct features remain constant for QNFs within the RN BH spacetime for each value of $\theta$: the value of $\mathbb{R}e \{ \Delta \omega \}$, where $\Delta \omega = \omega_{\ell+1} - \omega_{\ell}$, and $\vert \mathbb{I}m \{ \omega \} \vert$ in the large-$\ell$ limit.

\par For the Schwarzschild BH, we noted that an equidistant spacing $\mathbb{R}e \{ \Delta \omega \}$ emerged in the large-$\ell$ limit which matched that of the Schwarzschild Lyapunov exponent. In Table \ref{table:RNspacing}, we exploit the relationship between the orbital angular frequency and the Lyapunov exponent in the large multipolar limit, $\Omega_c \approx \overline{\Lambda}$ \cite{refGoebel1972,refCardosoLyapunov}, such that $\overline{\Lambda} = 1/b_c$ by eq. ({\ref{eq:wneg1}}) for $b_c$ defined in eq. ({\ref{eq:RNspecs}}), and thereby determine a unique RN Lyapunov exponent for each $\theta$. In so doing, we find a corresponding behaviour between $\overline{\Lambda}$ and $\mathbb{R}e \{ \Delta \omega \}$ for large $\ell$.

\par A further feature in the Schwarzschild case was a match between $\overline{\Lambda} /2$ and the value to which  $\mathbb{I}m \{ \omega \}$ tended as $\ell$ grew. For the RN BH, this relationship is only observed for lower values of $\theta$ to three decimal places. Furthermore, the consistency between the magnitude of the real part of the QNFs at large multipolar values is disrupted for larger values of $\theta$. 

\par The uniform potential once again produces QNFs that align closest to the Dirac results. However, it remains a fair intermediary between QNFs of integer and half-integer spin, and matches the behaviour of QNFs in the large-$\ell$ regime that we summarise in Table \ref{table:RNspacing}. We note, however, that the neat convergence observed in the Schwarzschild context for $\mathbb{R}e \{\omega \}$ across different spins for large values of $\ell$ is disrupted by the presence of $\theta$: though not a major effect, the discrepancy becomes apparent in integer-spin QNFs for larger $\theta$ values.  
\vfill

\begin{table}[t] 
\caption{\textit{Relationship between features of the QNFs in the large-$\ell$ limit and the Lyapunov exponent for all spins $s \in \{0,1/2,1,3/2,2\}$ and varying $\theta$ in RN BH spacetimes.}}
{\footnotesize{
\begin{tabular}{|C|C|C|C|C|C|C|}
\hline
&\theta = 0.0 & \theta = 0.2 & \theta = 0.4  & \theta = 0.6  & \theta = 0.8 & \theta = 1.0  \\
\hline
\mathbb{R}e \{ \Delta \omega \}& 0.1924 \pm 0.0001& 0.1937 \pm 0.0001 & 0.1979 \pm 0.0001 & 0.2058 \pm 0.0001 & 0.2199 \pm 0.0001 & 0.2499 \pm 0.0001  \\
\overline{\Lambda}  & 0.19245 & 0.19375 & 0.19790 &  0.20582 & 0.21997 & 0.2500 \\
\vert \mathbb{I}m \{ \omega \} \vert  & 0.0962 \pm 0.0001 & 0.0964 \pm 0.0001 & 0.0970 \pm 0.0001 & 0.0978 \pm 0.0001 & 0.0970 \pm 0.0001 & 0.0884 \pm 0.0001 \\
\overline{\Lambda} /2 & 0.09623 & 0.09688 & 0.09895 & 0.10291 & 0.10999 & 0.12500 \\
\hline
\end{tabular}
}}
\label{table:RNspacing}
\end{table}

\subsection{\label{subsec:SdS} QNF behaviour within the large-$\ell$ limit for SdS BHs}
\par The 4D SdS BH spacetime is markedly similar to its flat-space counterpart. From the metric function $f(r) = 1 - 2/r - \eta r^2/27$, where $\eta = 27 \lambda = 9 \Lambda$, we obtain
\begin{equation} \label{eq:SdSspecs}
r_c = 3 \;, \hspace{0.7cm}b_c = \sqrt{\frac{27}{1-\eta}} \; \;,  \hspace{0.7cm} \; \;
 \rho(r) = \left(1 - \frac{3}{r} \right) \sqrt{\frac{r+6}{(1-\eta) \; r}} \;.
\end{equation}

\par For the fields of integer spin, we follow the Schwarzschild example such that the effective potential used is that of eq. (\ref{eq:RWconciseHighD}), with $d=4$ and $\ell (\ell + 1) \rightarrow L^2 - 1/4$. For the spin-1/2 case, eq. ({\ref{eq:DODiracSchwarz}}) holds, albeit with the SdS metric function. The potential for the non-TT spin-3/2 field is fairly complicated, as we saw in eq. ({\ref{eq:RSSdSnontt}}). However, upon setting $d=4$, we find that
\begin{equation}
{\cal B} \;\big \vert_{d=4} = i \kappa \frac{\sqrt{f(r)}}{r} \; (z_{_{SdS}}+1) \hspace{0.5cm} \text{and} \hspace{0.5cm} {\cal D} \;\big \vert_{d=4} = -i \sqrt{\lambda f(r)} \; z_{_{SdS}} \;.
\end{equation}
\noindent Here, the $z_{_{SdS}}$ is the $z$ of eq. (\ref{eq:z}), which in 4D reduces to
\begin{equation}
z_{_{SdS}} \big \vert_{d=4} = \frac{1 - \left(1+ \frac{2\mu}{r} \right)}{\kappa^2 - \left(1- \frac{2\mu}{r} \right)} = \frac{1 - \left(1- \frac{2\mu}{r} \right)}{\left(1- \frac{2\mu}{r} \right) - \kappa^2} = z_s \;,
\end{equation}
\noindent where $z_s$ refers to eq. (\ref{eq:zs}). Thus, we observe that this behaviour associated with half-integer spin emerges once again, even if it is immersed in a greater equation of motion. In the 4D SdS case, however, this $z_{_{SdS}}$ retains the form of the asymptotically flat Schwarzschild BH, rather than receiving augmentation as in the RN case of eq. (\ref{eq:zrn}). Since
\begin{equation} 
\frac{\mathcal{D}}{\mathcal{B}}\; \bigg \vert_{d=4} =  \frac{2\sqrt{\lambda}}{\kappa(1- \kappa^2)} 
\end{equation}
\noindent is an $r$-independent expression, we observe that
\begin{equation}
\mathcal{F} \;\big \vert_{d=4} = f(r) \hspace{0.4cm} \text{and} \hspace{0.4cm} \mathcal{W} \;\big \vert_{d=4} = \sqrt{\mathcal{D}^2-\mathcal{B}^2} \;.
\end{equation}
\noindent We then perform a series expansion on $\mathcal{W}$ to ensure the boundary conditions specified by Ref. \cite{refDolanOttewill2009} are met:
\begin{equation}
\mathcal{W} \approx \frac{\kappa}{r}\sqrt{f(r)} + \frac{(16 + 9\lambda) f(r)}{8} \left( \frac{\kappa}{r}\sqrt{f(r)} \right)^{-1} ... \;.
\end{equation}
\noindent This expression is then substituted into eq. ({\ref{eq:SUSYpot}}) and the DO method is applied, with $\lambda \rightarrow \eta/27$ and $\kappa \rightarrow \bar{L}$.
\par In Ref. \cite{refDolanOttewill2009}, Dolan and Ottewill provide explicit expressions for eq. (\ref{eq:omegaseries}) up to $\mathcal{O} (L^{-4})$ for scalar perturbations in the SdS BH spacetime. In Table \ref{table:DOSdS}, we expand their work to include the QNFs for integer and half-integer spins up to $\mathcal{O}(L^{-6})$.

\begin{table}[!p]
\caption{\textit{The inverse multipolar expansions for the effective QNM potentials of spin $s$, where the scalar result is extended from Ref. \cite{refDolanOttewill2009} and subsequent expressions are new.}}
\begin{tabular}{| c | c |}
\hline
$s$ & $ b_c \sum^6_{k=-1}  \omega_k L^{-k} $ \\\hline
 & \textit{perturbations of integer spin} \\
\hline
& $L-\frac{i}{2} -\frac{61 \eta -7}{216 L} -i \frac{\left(-2005 \eta ^2+1868 \eta +137\right)}{7776 L^2} +\frac{750851 \eta ^3-1274856 \eta ^2+440043 \eta +5230}{2519424 L^3}$ \\
$0$& $+i \frac{\left(-135495065 \eta ^4+340616636 \eta ^3-243504102 \eta ^2+37791548 \eta +590983\right)}{362797056 L^4} $\\
& +$\frac{32505248 \eta ^5+1627761149 \eta ^4-3206256226 \eta ^3+1691631952 \eta ^2-161716742 \eta -9434549}{3265173504 L^5}$ \\
&$+i \frac{(1-\eta ) \left(34508780288 \eta ^5-21055361663 \eta ^4-34042960215 \eta ^3+20928906553 \eta ^2-1195508389 \eta -117456486\right)}{58773123072 L^6}$ \\
\hline
 & $L -\frac{i}{2} + \frac{(-65+11\eta)}{216L} + i\frac{5 (1-\eta ) (31 \eta +59)}{7776 L^2} -\frac{39709 \eta ^3-19848 \eta ^2-12363 \eta +71234}{2519424 L^3} $\\
$1$& $-i\frac{(1-\eta ) \left(494297 \eta ^3+181965 \eta ^2-366465 \eta -191699\right)}{22674816 L^4}$ \\
& $+\frac{(119468395 - 268062134 \eta + 261980704 \eta^2 -
 158224690 \eta^3 - 92371315 \eta^4 +
 111699872 \eta^5)}{3265173504 L^5} $\\
& $+i\frac{(1-\eta ) \left(3240662016 \eta ^5-200131567 \eta ^4-4986135335 \eta ^3+4932890217 \eta ^2-5358151637 \eta +2009486426\right)}{58773123072 L^6}$ \\
\hline
 & $L-\frac{i}{2}+\frac{-281+11 \eta }{216 L}+i\frac{(1-\eta ) (1591+155 \eta ))}{7776 L^2}-\frac{1420370+356997 \eta +151224 \eta ^2+39709 \eta ^3}{2519424 L^3} $\\
   & $ -i\frac{
   \left((1-\eta ) \left(-24667847+10941243 \eta +2695341 \eta ^2+494297 \eta ^3\right)\right)}{22674816 L^4} $\\
$2$   & $ +\frac{1036251595-130033958 \eta -6810863264 \eta ^2+2041863902 \eta ^3+562435853    \eta ^4+111699872 \eta ^5}{3265173504 L^5}$\\
   & $+i\frac{ \left((1-\eta ) \left(282174255002-262728531653 \eta -110347238007 \eta ^2+82283660137 \eta ^3+21001557905 \eta ^4+3240662016 \eta
   ^5\right)\right)}{58773123072 L^6}$\\
\hline
& \textit{the uniform potential} \\
\hline
& $ {\bar{L}}-\frac{i}{2}  +\frac{11 \eta -38}{216 {\bar{L}}} +i\frac{5 (1-\eta ) (31 \eta +59)}{7776 {\bar{L}}^2} -\frac{39709 \eta ^3-19848 \eta ^2+3675 \eta -3853}{2519424 {\bar{L}}^3} $ \\
& $-i\frac{ (1-\eta ) \left(988594 \eta ^3+363930 \eta ^2-464415 \eta +515945\right)}{45349632 {\bar{L}}^4} +\frac{111699872 \eta ^5-92371315 \eta
   ^4-128297944 \eta ^3+305515612 \eta ^2-349843298 \eta +150108427}{3265173504 {\bar{L}}^5} $\\
& $+ i\frac{ (1-\eta ) \left(3240662016 \eta ^5-200131567 \eta ^4-4024480547 \eta ^3+6837452937 \eta ^2-6191057384 \eta +901944887\right)}{58773123072 {\bar{L}}^6}$\\
\hline
& \textit{perturbations of half-integer spin} \\
\hline
& $ {\bar{L}}-\frac{i}{2}-\frac{11 (1-\eta )}{216 {\bar{L}}}-i\frac{ (-1+\eta ) (-29+155 \eta )}{7776 {\bar{L}}^2}-\frac{-3610+26517 \eta -62616 \eta ^2+39709 \eta ^3}{2519424 {\bar{L}}^3} $\\
 & $+\frac{(1-\eta )^{3/2} \left(-988594 i \sqrt{3} \sqrt{\frac{1}{1-\eta }} \eta ^3+3 \left(6316+231523 i \sqrt{3} \sqrt{\frac{1}{1-\eta }}\right) \eta ^2+\left(688836-311709 i \sqrt{3}
   \sqrt{\frac{1}{1-\eta }}\right) \eta +277684 i \sqrt{3} \sqrt{\frac{1}{1-\eta }}-2028732\right)}{45349632 \sqrt{3} {\bar{L}}^4} $\\
$1/2$  & $-\frac{(1-\eta )^{3/2} \left(68721752 i-10877291 \sqrt{\frac{3}{1-\eta }}-176643780 i \eta +14698199 \sqrt{\frac{3}{1-\eta }} \eta +45782556 i \eta ^2+6466191 \sqrt{\frac{3}{1-\eta }} \eta
   ^2\right)}{1632586752 \sqrt{3} {\bar{L}}^5} $ \\
& $-\frac{(1-\eta )^{3/2} \left(2578232 i \eta ^3+55980607 \sqrt{\frac{3}{1-\eta }} \eta ^3-55849936 \sqrt{\frac{3}{1-\eta }} \eta ^4\right)}{1632586752
   \sqrt{3} {\bar{L}}^5}$ \\
&$+\frac{(1-\eta )^{3/2} \left(2757868640+632909178 i \sqrt{\frac{3}{1-\eta }}-22621357508 \eta -5957507889 i \sqrt{\frac{3}{1-\eta }} \eta +34913622936 \eta ^2+7787617335 i
   \sqrt{\frac{3}{1-\eta }} \eta ^2\right)}{176319369216 \sqrt{3} {\bar{L}}^6}$\\
&$-\frac{(1-\eta )^{3/2} \left(8255611364 \eta ^3+1837367991 i \sqrt{\frac{3}{1-\eta }} \eta ^3+1555478288 \eta
   ^4-13014760527 i \sqrt{\frac{3}{1-\eta }} \eta ^4+9721986048 i \sqrt{\frac{3}{1-\eta }} \eta ^5\right)}{176319369216 \sqrt{3} {\bar{L}}^6}$\\
\hline
\end{tabular}
\label{table:DOSdS}
\end{table}

\begin{table}[t]
\caption*{}
\begin{tabular}{| c | c |}
\hline
 & ${\bar{L}}+\frac{-9795520512 i+9795520512 i \eta }{19591041024 (1-\eta )}+\frac{-16507266048+997691904 \eta }{19591041024 {\bar{L}}} +\frac{2920012416 i-5449514112 i \eta +2138990976 i \eta ^2+390510720 i \eta ^3}{(1-\eta ) 19591041024 {\bar{L}}^2} $\\
& $+\frac{2513531735}{19591041024 \sqrt{3-3 \eta } {\bar{L}}^2}  +\frac{-9795520512 i+9795520512 i \eta ^5}{19591041024 \sqrt{3-3 \eta } {\bar{L}}^3} +\frac{473379552=-1217744928 \eta -732499200 \eta ^2-308777184 \eta ^3}{19591041024
   {\bar{L}}^3}$ \\
& $+\frac{8658485712 i-21708103248 i \eta +15835827024 i \eta ^2-1608359760 i \eta ^3-750777120 i \eta
   ^4-427072608 i \eta ^5}{19591041024 (1-\eta ) {\bar{L}}^4} $ \\
 $3/2$ & $+\frac{39169484928-18966223872 \eta }{19591041024 \sqrt{3-3 \eta } {\bar{L}}^4}+\frac{5063254530+9042938484 \eta -19353693144 \eta ^2+4363840560
   \eta ^3+2065000782 \eta ^4+670199232 \eta ^5}{19591041024 {\bar{L}}^5}$ \\
& $+\frac{14574447936 i-70857540288 i \eta +88207553664 i \eta ^2-31924461312 i \eta ^3}{19591041024 \sqrt{3-3 \eta }
   {\bar{L}}^5} +\frac{2484334683 i \eta ^6+1080220672 i \eta ^7}{19591041024 (1-\eta ) {\bar{L}}^6}$ \\
& $+\frac{17935074621 i-54280336946 i \eta +31035646320 i \eta ^2+37626780033 i \eta ^3-38990726240 i \eta ^4+3109006857 i \eta ^5}{19591041024
   (1-\eta ) {\bar{L}}^6} $ \\
& $+\frac{11886526656+16404841872 \eta -9940944672 \eta ^2-2367512928 \eta ^3}{19591041024 \sqrt{3-3 \eta } {\bar{L}}^6}$  \\
\hline
\end{tabular}
\end{table}

\par In this section, we calculate the QNFs for the 4D SdS BH using these expressions from Table \ref{table:DOSdS} for each of the fields of spin $s \in \{0,1/2,1,3/2,2\}$. We address QNFs of integer and half-integer spin separately. The QNFs of the SdS BH within the eikonal regime follow the same trends observed for the Schwarzschild and RN BH spacetimes: a steadily increasing real part and a constant imaginary part (emergent for each value of $\Lambda$ and related to the associated Lyapunov exponent), as well as a decrease (increase) in the real part of the QNFs of integer spin (half-integer spin) for increasing $s$ and fixed $\ell$. 
\par Moreover, an equidistant spacing between subsequent QNFs emerges for each value of $\Lambda$, shown in Table \ref{table:SdSspacing} to be equivalent to the Lyapunov exponent as defined in eq. (\ref{eq:LyapunovSdS}), much like in the Schwarzschild case. Since $\Omega_c \approx \overline{\Lambda}$ in the large-$\ell$ regime \cite{refGoebel1972,refCardosoLyapunov}, this spacing is equivalent also to the orbital angular frequency. Furthermore, the $\Lambda = 0$ column matches its asymptotically flat-space counterpart, as expected. However, as the value of $\Lambda$ increases, the growth of both the real and the imaginary part is hindered: $\mathbb{R}e \{ \Delta \omega \}$ decreases in magnitude and the imaginary part converges to a smaller value. From this, we might surmise that the presence of a positive spacetime curvature suppresses the influence of angular momentum on the QNF.

\par We first list the low-lying results and compare these to extant QNF expressions, to further the validation of the method as initiated in Ref. \cite{refDolanOttewill2009} (see appendix \ref{app:SdS}, where we find excellent agreement with Ref. \cite{refZhidenko2004}). We also verify our large-$\ell$ results with the WKB method of Ref. \cite{Konoplya2003} and the PT approximation of Refs. \cite{PoschTellerMethod,refFerrariMashhoon}, and determine that the results are highly consistent.

\subsubsection{\label{subsec:SdSinteger} QNFs of integer spin within 4D SdS BH spacetimes}

\begingroup
\begin{table}[t]
\caption{\textit{QNFs of integer spin for a 4D SdS BH calculated via the DO method to order $\mathcal{O}(L^{-6})$ for spin-0, spin-1, and spin-2 fields.}}
\begin{tabular}{|C|C|C|C|C|C|}
\hline
field & \ell & \omega \; (\Lambda = 0.00) & \omega \; (\Lambda = 0.04)&  \omega \; (\Lambda = 0.08) & \omega \; (\Lambda = 0.10) \\
\hline
		& 10 & 2.0213-0.0963 i & 1.6156-0.0771 i & 1.0676-0.0510 i & 0.6377-0.0304 i \\
		& 20 & 3.9455-0.0962 i & 3.1557-0.0770 i & 2.0868-0.0509 i & 1.2469-0.0304 i \\
 scalar & 40 & 7.7944-0.0962 i & 6.2351-0.0770 i & 4.1239-0.0509 i & 2.4644-0.0304 i \\
		& 60 & 11.6433-0.0962 i & 9.3144-0.0770 i & 6.1607-0.0509 i & 3.6817-0.0304 i \\
		& 80 & 15.4920-0.0962 i & 12.3937-0.0770 i & 8.1975-0.0509 i & 4.8989-0.0304 i \\
		& 100 & 19.3410-0.0962 i & 15.4730-0.0770 i & 10.2342-0.0509 i & 6.1161-0.0304 i \\
\hline
		&10 & 2.0152-0.0962 i & 1.6124-0.0769 i & 1.0667-0.0509 i & 0.6375-0.0304 i \\
		& 20 & 3.9424-0.0962 i & 3.1541-0.0770 i & 2.0863-0.0509 i & 1.2468-0.0304 i \\
 EM 		&40 & 7.7928-0.0962 i & 6.2343-0.0770 i & 4.1237-0.0509 i & 2.4644-0.0304 i \\
		& 60 & 11.6423-0.0962 i & 9.3139-0.0770 i & 6.1606-0.0509 i & 3.6817-0.0304 i \\
		& 80 & 15.4920-0.0962 i & 12.3932-0.0770 i & 8.1974-0.0509 i & 4.8989-0.0304 i \\
		& 100 & 19.3410-0.0962 i & 15.4730-0.0770 i & 10.2342-0.0509 i & 6.1161-0.0304 i \\
\hline
		& 10 & 1.9968-0.0958 i & 1.5977-0.0768 i & 1.0569-0.0509 i & 0.6317-0.0304 i \\
		& 20 & 3.9330-0.0961 i & 3.1465-0.0769 i & 2.0813-0.0509 i & 1.2439-0.0304 i \\
 grav.	& 40 & 7.7880-0.0962 i & 6.2305-0.0770 i & 4.1211-0.0509 i & 2.4629-0.0304 i \\
		& 60 & 11.6391-0.0962 i & 9.3113-0.0770 i & 6.1589-0.0509 i & 3.6807-0.0304 i \\
		& 80 & 15.4890-0.0962 i & 12.3913-0.0770 i & 8.1961-0.0509 i & 4.8981-0.0304 i \\
		& 100 & 19.3390-0.0962 i & 15.4710-0.0770 i & 10.2331-0.0509 i & 6.1155-0.0304 i \\
 \hline
\end{tabular}
\label{table:SdSIntegerLarge}
\end{table}
\endgroup

\begin{table}[t]
\caption{\textit{QNFs of half-integer spin and the ``uniform" potential for a 4D SdS BH calculated via the DO method to order $\mathcal{O}(L^{-6})$.}}
\begin{tabular}{|C|C|C|C|C|C|}
\hline
field & \ell & \omega \; (\Lambda = 0.00) & \omega \; (\Lambda = 0.04)&  \omega \; (\Lambda = 0.08) & \omega \; (\Lambda = 0.10) \\
\hline
 &10 & 2.1139-0.0962 i & 1.6914-0.0769 i & 1.1189-0.0509 i & 0.6687-0.0304 i \\
 &20 & 4.0398-0.0962 i & 3.2320-0.0770 i & 2.1379-0.0509 i & 1.2776-0.0304 i \\
 uniform&40 & 7.8896-0.0962 i & 6.3118-0.0770 i & 4.1749-0.0509 i & 2.4950-0.0304 i \\
 &60 & 11.7389-0.0962 i & 9.3912-0.0770 i & 6.2117-0.0509 i & 3.7122-0.0304 i \\
 &80 & 15.5880-0.0962 i & 12.4705-0.0770 i & 8.2485-0.0509 i & 4.9294-0.0304 i \\
 &100 & 19.4370-0.0962 i & 15.5500-0.0770 i & 10.2852-0.0509 i & 6.1466-0.0304 i \\
 \hline
		& 10 & 2.1161-0.0962 i & 1.6931-0.0770 i & 1.1201-0.0509 i & 0.6694-0.0304 i \\
		& 20 & 4.0410-0.0962 i & 3.2329-0.0770 i & 2.1385-0.0509 i & 1.2780-0.0304 i \\
Dirac	& 40 & 7.8902-0.0962 i & 6.3122-0.0770 i & 4.1752-0.0509 i & 2.4952-0.0304 i \\
		& 60 & 11.7393-0.0962 i & 9.3915-0.0770 i & 6.2119-0.0509 i & 3.7123-0.0304 i \\
		& 80 & 15.5880-0.0962 i & 12.4707-0.0770 i & 8.2486-0.0509 i & 4.9295-0.0304 i \\
		& 100 & 19.4370-0.0962 i & 15.5500-0.0770 i & 10.2853-0.0509 i & 6.1467-0.0304 i \\
\hline
		& 10 & 2.2964-0.0961 i & 1.8373-0.0769 i & 1.2153-0.0509 i & 0.7263-0.0304 i \\
	& 20 & 4.2267-0.0962 i & 3.3815-0.0770 i & 2.2367-0.0509 i & 1.3367-0.0304 i \\
RS & 40 & 8.0791-0.0962 i & 6.4633-0.0770 i & 4.2751-0.0509 i & 2.5549-0.0304 i \\
	& 60 & 11.9293-0.0962 i & 9.5435-0.0770 i & 6.3125-0.0509 i & 3.7724-0.0304 i \\
	& 80 & 15.7790-0.0962 i & 12.6232-0.0770 i & 8.3495-0.0509 i & 4.9898-0.0304 i \\
	& 100 & 19.6280-0.0962 i & 15.7030-0.0770 i & 10.3864-0.0509 i & 6.2070-0.0304 i \\
 \hline
\end{tabular}
\label{table:SdSHalfIntegerLarge}
\end{table}

\par In keeping with the behaviour of the QNFs within the Schwarzschild and RN BH spacetimes, the magnitude of the real part of the QNF decreases for increasing spin for SdS BH spacetimes. For larger values of $\ell$, we observe that $\mathbb{R}e \{ \omega \}$ of different spins become uniform. If we compare the large-$\ell$ QNFs of scalar and electromagnetic fields, the QNFs match exceedingly well. Thus, we observe that the spin of the oscillation does not influence QNF behaviour in the large-$\ell$ limit. This is further supported by the consistent values obtained for the spacing between subsequent QNFs, as shown in Table \ref{table:SdSspacing}.

\par While the same trends apply for the gravitational fields, we note that there remains a slight discrepancy at $\ell = 100$ between the QNFs for the gravitational and electromagnetic perturbations. In other words, the effect of spin is not as quickly negated for the spn-2 fields when compared with the other integer fields studied.
\vfill

\subsubsection{\label{subsec:SdSHalfInteger} QNFs of half-integer spin within 4D SdS BH spacetimes}
\par Despite the approximations introduced, the spin-1/2 and spin-3/2 QNFs derived with the DO method remain in excellent agreement with the WKB and PT results we calculate (where low-lying points of comparison are recorded in appendix \ref{app:SdS}).

\par The QNFs of half-integer spin reflect the behaviour of the integer fields, with an equivalent suppressive influence observed for increasing values of $\Lambda$. Though the magnitudes of $\mathbb{R}e \{ \omega \}$ for the Dirac and Rarita-Schwinger fields do not match exactly for $\ell = 100$, the discrepancy between the two remains small such that the role of spin becomes demonstrably diminished in the large-$\ell$ limit. 

\par As we have seen for the Schwarzschild and RN BHs, the Dirac and the uniform QNFs correlate best. However, the QNFs associated with this uniform potential carry the same characteristics observed for all fields within the large multipolar limit for each $\Lambda$, most notably the emergence of an equidistant spacing in the real part and a constant imaginary part, both of which relate to the Lyapunov exponent and the orbital angular frequency in the large-$\ell$ regime.  

\begin{table}[t]
\caption{\textit{Relationship between features of the QNFs in the large-$\ell$ limit and the Lyapunov exponent for all spins $s \in \{0,1/2,1,3/2,2\}$ and varying $\Lambda$ in 4D SdS BHs.}}
{\footnotesize{
\begin{tabular}{|C|C|C|C|C|C|C|}
\hline
&\Lambda = 0.00 & \Lambda = 0.02 & \Lambda = 0.04  & \Lambda = 0.06  & \Lambda = 0.08 & \Lambda = 0.10  \\
\hline
\mathbb{R}e \{ \Delta \omega \}& 0.1924 \pm 0.0001& 0.1743 \pm 0.0001 & 0.1540 \pm 0.0001 & 0.1305 \pm 0.0001 & 0.1018 \pm 0.0001 & 0.0608 \pm 0.0001  \\
\overline{\Lambda}  & 0.1925 & 0.1743 & 0.1540 &  0.1305 & 0.1018 & 0.06086 \\
\vert \mathbb{I}m \{ \omega \} \vert  & 0.0962 \pm 0.0001 & 0.0871 \pm 0.0001 & 0.0770 \pm 0.0001 & 0.0653 \pm 0.0001 & 0.05092 \pm 0.0001 & 0.0304 \pm 0.0001\\
\overline{\Lambda} /2 & 0.0962 & 0.0871 & 0.0770 & 0.0653 & 0.05092 & 0.0304 \\
\hline
\end{tabular}
}}
\label{table:SdSspacing}
\end{table}

\par In Table \ref{table:SdSspacing}, the quantitative commonalities  that emerge irrespective of the spin of the QNF are recorded. Here, we calculate the Lyapunov exponent using eq. (\ref{eq:LyapunovSdS}); since we are in the large multipolar limit, we know that the orbital angular frequency is equivalent to this quantity \cite{refGoebel1972,refCardosoLyapunov}. The relationships between the Lyapunov exponent, as well as $\mathbb{R}e \{ \Delta \omega \}$ and $\vert \mathbb{I}m \{ \omega \} \vert$ in the large-$\ell$ limit, are as in the Schwarzschild case, such that the Lyapunov exponent can be extracted directly from the inter-QNF spacing and the constant imaginary part of the QNF associated with each value of $\Lambda$. As $\Lambda$ increases, the Lyapunov exponent and all related quantities decrease.

%
%
\section{\label{sec:conc}Conclusions}

\par Within this work, we have performed a review of the extant expressions for the effective QNM potentials in the literature associated with perturbing fields of integer and half-integer spin in stationary, spherically-symmetric BH spacetimes of $d \geq 4$. Through this systematic analysis, we have determined that the application of the large multipolar limit reduces these effective potentials to a common form, irrespective of the spin of the field and regardless of the nuances of the BH spacetime (i.e. BH mass, BH charge, etc.), provided $\lambda \geq 0$. For $\lambda < 0$, the asymptotic behaviour of $r$ affects the final outcome. However, the effect is uniform for all AdS cases studied here, such that $V_{eff} \rightarrow constant$ is universal for our choice of AdS boundary conditions.

\par Since many of the mathematical techniques historically applied to QNM problems exhibit greater accuracy when $\ell \gg n$, this uniformity of expression suggests that eq. ({\ref{eq:bigL}}) alone is consistently sufficient to produce reliable results within Minkowski and dS spacetimes; for AdS spacetimes, the behaviour exhibited in eq. ({\ref{eq:bigLAdS}}) should suffice. Furthermore, if we were to automate a QNM calculation procedure, this observation implies that dimensionality and angular momentum serve as the primary factors to consider. As such, with the generalised expressions of eqs. ({\ref{eq:bigL}}) and ({\ref{eq:bigLAdS}}) in place, we are in a better position to compare various QNM problem-solving methods in order to gauge their  relative accuracy within the large multipolar limit, and to exploit this consistency in the development of new mathematical techniques and possibly machine-learning algorithms.

\par To validate the behaviour observed in our analytical investigation of the large-$\ell$ regime, we have engaged in a numerical study of QNFs for increasing values of $\ell$ via Dolan and Ottewill's recently developed inverse multipolar expansion method, with a focus on QNFs of spin $s \in \{ 0,1/2,1,3/2,2\}$ in Schwarzschild, RN, and SdS spacetimes. In section \ref{sec:num}, we have provided a full description of the physical origins of the method and the manner in which it is applied; we have addressed the necessary ansatz and associated components required for QNF computation in each spacetime and, where feasible, explicitly recorded the series expansion with which the QNFs are to computed for each field. Except for gravitational and electromagnetic QNFs of the RN BH spacetime with $\theta < 1$, all results have been carried to orders of $\mathcal{O} (L^{-6})$, reflecting a marked improvement on several extant attempts at pursuing the DO method in the literature. At this order, we found excellent agreement between the QNFs we computed using the DO method and those obtained through the 6th-order WKB and PT methods, particularly for larger values of $\ell$. We anticipate better agreement for expansions at higher orders of $L^{-k}$. 

\par Irrespective of the BH context and the spin of the field, we observed that the magnitude of the QNFs decreased (increased) with $s$ for QNFs of integer (half-integer) spin for fixed $\ell$ $-$ an effect more pronounced in the lower-$\ell$ regime. In accordance with our analytical results in the large multipolar limit, the magnitude of the QNFs for different values of $s$ began to converge for large $\ell$. This was noted explicitly in the Schwarzschild and SdS cases (for each value of the cosmological constant, with a distinct value of integer-spin QNFs); the presence of $\theta$ slightly offset this consistency. Furthermore, we observed that $\mathbb{R}e \{ \omega \}$ increased with increasing $\theta$ while both $\mathbb{R}e \{ \omega \}$ and $\mathbb{I}m \{ \omega \}$ decreased with increasing $\lambda$ $-$ an effect more pronounced in the large-$\ell$ regime.   

\par From these observations, we may surmise that the nature of the BH parameters is of greater influence than the nuances of the perturbing field within the large-$\ell$ regime. The effect of an increase in $\theta$ manifests as an increase in $\mathbb{R}e \{\omega \}$ and a slight decrease in $\mathbb{I}m \{ \omega \}$ (please see Tables \ref{table:RNIntegerLarge} and \ref{table:RNHalfIntegerLarge}). This implies an increase in oscillation frequency and a decrease in damping, thereby demonstrating that BH charge increases QNF energy. In contrast, $\Lambda$ suppresses QNF growth for both $\mathbb{R}e \{\omega \}$ and  $\mathbb{I}m \{ \omega \}$. This effect appears more pronounced than that of the BH charge, with relatively substantial differences in QNF magnitude seen as $\Lambda$ was increased from $0.00$ to $0.1$ for large $\ell$ (please see Tables \ref{table:SdSIntegerLarge} and \ref{table:SdSHalfIntegerLarge}). This suggests that the QNF dependence on the cosmological constant is particularly significant. Although we maintained a constant BH mass throughout our investigation, it would be interesting to observe the effect of BH mass on QNF magnitudes.

\par Based on known behaviours in the literature (Refs. \cite{refGoebel1972,refCardosoLyapunov,refZhidenko2004}), we expected the imaginary part of the gravitational QNF to converge to a constant that $-$ in the Schwarzschild and SdS BH spacetimes $-$ would match the Lyapunov exponent. We found that this relationship between $\mathbb{I}m \{ \omega \}$ and $\overline{\Lambda}$ for $\ell \rightarrow \infty$ applied to QNFs of all spins studied here for Schwarzschild, SdS, and RN BH spacetimes of sufficiently low $\theta$. An unexpected observation was the emergence of a constant spacing between successive values of $\mathbb{R}e \{ \omega \}$ within the large-$\ell$ regime that matched precisely with this Lyapunov exponent. We suggest that this is related to the fact that $\overline{\Lambda} \approx \Omega_c$ within the large multipolar limit.  

\par Thus, we have demonstrated that the large multipolar regime offers a number of physical insights regarding the behaviour of QNFs within various spacetimes, for which further study is warranted. Additional investigations into the DO method is also required, such as a full-scale analysis of the form we presented here focused instead on the QNM wavefunction. Further comparative studies between the DO method at higher orders of $L^{-k}$ and more recently constructed methods with demonstrably enhanced accuracy (such as the improved semianalytic approach of Refs. \cite{Matyjasek2017,Matyjasek2019}) are of particular interest. Such studies would benefit from the inclusion of an explicit error analysis, which was not considered in this work. Finally, we note that extensions of the method to higher dimensions and AdS spacetimes are highly desirable.

%
%
\section*{Acknowledgments}

HTC is supported in part by the Ministry of Science and Technology, Taiwan, under the Grants No. MOST108-2112-M-032-002 and MOST109-2112-M-032-007. AC is supported by the National Institute for Theoretical Physics, South Africa. ASC is supported in part by the National Research Foundation of South Africa.

\appendix
\section{\label{app}Validation of the DO method for low-lying QNFs in 4D BH spacetimes}
\par In Ref. \cite{refDolanOttewill2009}, Dolan and Ottewill explicitly compared their results for the gravitational QNFs within a Schwarzschild BH spacetime with extant results in the literature. They observed that the excellent agreement they found for these perturbations extended to those of integer spin for the Schwarzschild case. Expansions to higher orders of $L^{-k}$ corresponded to improved agreement. Here, we continue their validation of the DO method in the low-$\ell$ regime: we supply sources of comparison from the literature for half-integer QNFs within the Schwarzschild BH spacetime, and for QNFs of all spins studied in this paper within the RN and SdS BH spacetimes. Where appropriate, we provide results we calculate using the 6th-order WKB and PT methods, produced with the aid of the Mathematica notebooks made available in Refs. \cite{Konoplya2003} and \cite{refZhidenko2004}.

\subsection{QNFs in Schwarzschild BH spacetimes \label{app:Schwarz}}
\par The DO results of Tables \ref{table:SchwarzComp1} and \ref{table:SchwarzComp2} are computed using the $\sum_k \omega_k L^{-k} $ expansions to $\mathcal{O}(L^{-6})$, extracted from the $b_c \sum^6_{k=-1} \omega_k L^{-k}$ expansions of Table \ref{table:DOschwarz}. The positive imaginary part in Shu and Shen's results \cite{refShuShen} is a consequence of their choice in temporal dependence.

\par In Table \ref{table:SchwarzComp1}, the columns labelled $\omega_{SS}$ and $\omega_{Cho}$ showcase results from Shu and Shen \cite{refShuShen} and Cho \cite{refDirac03}, respectively, derived via the 3rd-order WKB. The results of the remaining columns are calculated by us using the 6th-order WKB, PT, and DO methods. 

\par In Table \ref{table:SchwarzComp2}, we again use the 3rd-order WKB results of Ref. \cite{refShuShen}. The 3rd-order WKB, 6th-order WKB, and improved AIM results are taken from Ref. \cite{refRSSchwarz16}. We calculate the final DO column to $\mathcal{O}(L^{-6})$. 
\vfill

\begin{table}[h!]
\centering
\caption{\textit{Spin-1/2 QNFs of Schwarzschild BHs from Refs. \cite{refShuShen,refDirac03} and computed with the 6th-order WKB, PT approximation, and DO expansion from Table \ref{table:DOschwarz}.}}
\begin{tabular}{| C | L | L | L | L | L |}
\hline
\ell & \hspace{0.8cm} \omega_{SS} \; \text{\cite{refShuShen}} & \hspace{0.6cm} \omega_{Cho}   \; \text{\cite{refDirac03}} & \hspace{0.3cm} \omega_{calc} \; \text{(WKB)} & \hspace{0.6cm} \omega_{calc} \; \text{(PT)} & \hspace{0.6cm} \omega_{calc} \; \text{(DO)} \\
\hline
1 & 0.3786+0.0965i & 0.379-0.097i & 0.3801-0.0964i & 0.3855-0.0991i & 0.3800-0.0964i \\
2 & 0.5737+0.0963i & 0.574-0.096i & 0.5741-0.0963i  & 0.5779-0.0975i & 0.5741-0.0963i \\
3 & 0.7672+0.0963i & 0.7670-0.096i & 0.7674-0.0963i  & 0.7702-0.0969i & 0.7674-0.0963i \\
4  & 0.9602+0.0963i & 0.960-0.096i & 0.9603-0.0963i  & 0.9625-0.0963i & 0.9603 - 0.0963i \\
  \hline
\end{tabular}
\label{table:SchwarzComp1}
\end{table}

\begin{table}[h!]
\caption{\textit{Spin-3/2 QNFs of Schwarzschild BHs from Refs. \cite{refShuShen,refRSSchwarz16} based on the WKB and AIM, and computed with the DO expansion from Table \ref{table:DOschwarz}.}}
\begin{tabular}{| C | L | L | L | L | L |}
\hline
\ell & \hspace{0.8cm} \omega_{SS} \; \text{\cite{refShuShen}} & \hspace{0.8cm} \omega_{3rd} \; \text{\cite{refRSSchwarz16}} & \hspace{0.8cm} \omega_{6th} \; \text{\cite{refRSSchwarz16}} & \hspace{0.65cm} \omega_{AIM} \; \text{\cite{refRSSchwarz16}} & \hspace{0.6cm} \omega_{calc} \; \text{(DO)}  
\\
\hline
2 & 0.7346+0.0949i & 0.7346-0.0949i & 0.7348-0.0949i & 0.7347-0.0948i & 0.7348-0.0949i\\
3 & 0.9343+0.0954i & 0.9343-0.0954i & 0.9344-0.0954i & 0.9343-0.0953i & 0.9345 -0.0954i\\
4 & 1.1315+0.0956i & 1.1315-0.0956i & 1.1315-0.0956i & 1.1315-0.0956i & 1.1315-0.0956i\\
\hline
\end{tabular}
\label{table:SchwarzComp2}
\end{table}

\subsection{QNFs in Reissner-Nordstr{\"o}m BH spacetimes \label{app:RN}}
\par For the RN BH spacetimes, the DO method extends to order $\mathcal{O}(L^{-6})$ for QNFs of spin $s=0,1/2,3/2$ without incident. However, for QNFs associated with electromagnetic and gravitational perturbations, complications arise. This can be seen in the following comparisons with the literature.  

\par In Table \ref{table:RNScalarLow}, we compare the QNFs we produce via the DO method to $\mathcal{O}(L^{-6})$ with results from Ref. \cite{refFernandoCorrea} calculated using the 6th-order WKB method. We include results we calculated using the WKB method and PT approximation to improve our comparison. 
\par For lower values of $\theta$, we find a closer agreement between the WKB and DO results than the WKB and the PT for the spin-0 QNFs, such that the DO method appears more accurate than the PT under these conditions.

\par For the electromagnetic case, we calculate the expansions up to $\mathcal{O}(L^{-4})$ for $\theta <1$ and to $\mathcal{O}(L^{-6})$ for $\theta =1$. We compare our results in Table \ref{table:RNEMLow} with those listed in Ref. \cite{refBHWKB3}, which includes the QNFs determined in Ref. \cite{refGunter1980RN} by Gunter. Note that where we use $\theta = 1$ for the DO method, these references make use of $\theta = 0.99$ to accommodate the limitations of their chosen methodologies. In our own calculations for the QNFs of the RN BH spacetime using the 6th-order WKB and PT approximation, we have seen the methods fail as $\theta \rightarrow 1$.

\par For the low-lying gravitational QNFs, we compare our results with those listed in Ref. \cite{refBHWKB3}, where Kokkotas and Schultz employ the 3rd-order WKB method and compare their results against those Gunter obtained numerically in Ref. \cite{refGunter1980RN}. In Table \ref{table:RNGravLow}, we observe that Gunter's results match or overtake the DO for lower values of $\theta$, particularly for $\mathbb{I}m \{ \omega \}$. However, the DO method becomes demonstrably more accurate for larger $\ell$.

\par Relative to their integer-spin counterparts, there are few reviews of spin-1/2 QNFs for RN BHs, particularly using the formalism we apply. We compare the DO results we calculate to order $\mathcal{O}(L^{-6})$ and those of Ref. \cite{refJing2003} using the PT approximation in Table \ref{table:RNDiracLow}. The results match best for $\theta = 0.4$; the DO QNFs show improvement as $\ell$ increases.

\par In Table \ref{table:RNRSLow}, we compare the DO method for spin-3/2 QNFs to order $\mathcal{O}(L^{-6})$ with results obtained via the 6th-order WKB and the improved AIM, as recorded in Ref. \cite{refRSRN18}. Though the DO method demonstrates relatively poor agreement for results of low multipolar number, particularly for $\mathbb{I}m \{ \omega \}$, it becomes more reliable as $\ell$ and $\theta$ increase.

\begin{table}[t]
\caption{\textit{Spin-0 QNFs for 4D RN BHs calculated using the DO method and compared with the 6th-order WKB from Ref. \cite{refFernandoCorrea}
and the WKB and PT results we calculate.}}
\begin{tabular}{|C|C|C|C|C|}
\hline
\ell & \omega \; (\theta = 0.2) & \omega \; (\theta = 0.4) & \omega \; (\theta = 0.6) & \omega \; (\theta = 0.8)  \\
\hline
2 \; (DO) & 0.4876 - 0.0971i & 0.5001 - 0.0978i & 0.5245 - 0.0989i &
  0.6078 - 0.0973i \\
2 \; \text {\cite{refFernandoCorrea}}  & 0.4869-0.09697i & 0.4974-0.09756i & 0.5174-0.09833i & 0.5531-0.09834i \\
2 \; (PT) & 0.4913-0.0982 i & 0.5041-0.0989 i & 0.5288-0.0998 i & 0.5747-0.1001 i \\
\hline
3 \; (DO) & 0.6804 - 0.0967i & 0.6970 - 0.0974i & 0.7277 - 0.0984i & 0.8318 - 0.0967i \\
3 \; (WKB) & 0.6805 - 0.0967i &  0.6967 - 0.0974i & 0.7281 - 0.0983i &  0.7853 - 0.0985i \\
3 \; (PT) & 0.6832 - 0.0973i & 0.6994 - 0.0980i & 0.7306 - 0.0989i &  0.7876 - 0.0990i \\
\hline
\end{tabular}
\label{table:RNScalarLow}
\end{table}

\begin{table}[t]
\caption{\textit{Spin-1 QNFs in the 4D RN BH spacetime calculated using the DO method and compared with the 3rd-order WKB results of Ref. \cite{refBHWKB3} and numerical results of Ref. \cite{refGunter1980RN}.}}
\begin{tabular}{|C|C|C|C|C|}
\hline
\ell & \omega \; (\theta=0.0) & \omega \; (\theta=0.4) &  \omega \; (\theta=0.8) &\omega  \;(\theta \approx 1)\\
\hline
2 \; (DO) & 0.4576 - 0.09501i &  0.4751 - 0.0966i &
0.5702 - 0.09899i & 0.7043 - 0.0859i \\
2  \; \text{\cite{refBHWKB3}} &	0.4571-0.0951i	&	0.4795-0.0965i		&	0.5698-0.0990i & 	0.6925-0.0892i \\
2 \; \text{\cite{refGunter1980RN}}  &	0.4576-0.0950i	&	0.4799 -0.0964i		&	0.5701-0.0991i	& 0.6928-0.0886i \\
\hline
3 \; (DO) & 0.6569 - 0.0956i &  0.6864 - 0.0970i & 0.8029 - 0.0991i & 0.9658 - 0.0870i\\
3 \; \text{\cite{refBHWKB3}}  & 0.6567-0.0956i & 0.6871-0.0970i &			0.8027-0.0991i & 0.9519-0.0889i \\
3 \; \text{\cite{refGunter1980RN}} & 0.656-0.0956i	&	0.6873 -0.0970i	 & 0.8028-0.0991i &	0.9521-0.0893i \\
\hline
4 \; (DO) & 0.8531 - 0.0959i &  0.8908 - 0.0972i & 1.0304 - 0.0990i & 1.2226 - 0.0875i \\
4 \; \text{\cite{refBHWKB3}}  & 0.8530-0.0959i	 & 0.8909-0.0972i &		1.0303-0.0991i & 1.2066-0.0891i \\
4 \; \text{\cite{refGunter1980RN}} &	0.8531-0.0959i &		0.891-0.0972i &	 1.0304-0.0990i	& 1.2067-0.0898i \\
\hline
 5 \; \text{\cite{refBHWKB3}}  & 1.0382-0.0960i	& 1.0929-0.0972i &		1.2554-0.0989i &	 1.4588-0.0896i \\
5 \; (DO) &1.0479 - 0.0960i & 1.0928 - 0.0972i & 1.2554 - 0.0989i &  1.4772 - 0.0878i \\
 5 \;\text{\cite{refGunter1980RN}}  & 1.0459-0.0960i &		1.0929-0.0974i &	 1.2554-0.0947i & 1.4589-0.0898i \\
\hline
\end{tabular}
\label{table:RNEMLow}
\end{table}

\begin{table}[t]
\caption{\textit{Spin-2 QNFs for 4D RN BHs calculated using the DO method and compared with the 3rd-order WKB results of Ref. \cite{refBHWKB3} and numerical results of Ref. \cite{refGunter1980RN}.}}
\begin{tabular}{|C|C|C|C|C|}
\hline
\ell & \omega \; (\theta=0.0) & \omega \; (\theta=0.4) &  \omega \; (\theta=0.8) &\omega  \;(\theta \approx 1)\\
\hline
2 \; (DO) &  0.3736 - 0.0887i & 0.3836 - 0.0889i &  0.4013 - 0.0892i &  0.4313 - 0.0833i \\
2 \; \text{\cite{refBHWKB3}} & 0.3732-0.0892i &	 0.3779-0.0896i & 0.4005-0.0898i &	0.4283-0.0853i \\
2 \; \text{\cite{refGunter1980RN}} & 0.3737-0.0890i &	 0.3784-0.0894i & 0.4012-0.0896i & 	0.4293-0.0843i \\
\hline
3 \; (DO) & 0.5994 - 0.0927i &  0.6080 - 0.09300i & 0.6476 - 0.0931i & 0.7043 - 0.0860 i \\
3 \; \text{\cite{refBHWKB3}} & 0.5993-0.0927i & 0.6069 -0.0931i & 0.6473-0.0931i &	0.6998 -0.0871i  \\
3 \; \text{\cite{refGunter1980RN}} & 0.5994-0.0927i &	 0.6071 - 0.0931i & 0.6476-0.0931i &	0.7001 -0.0870i \\
\hline
4 \; (DO) & 0.8092 - 0.09416i & 0.8204 - 0.0945i & 0.8806 - 0.09466i & 0.9658 - 0.0870i \\
4 \; \text{\cite{refBHWKB3}} & 0.8091-0.0942i &	 0.8201-0.0945i & 0.8805-0.0947i &	0.9591-0.08800i \\
4 \; \text{\cite{refGunter1980RN}} & 0.8092-0.0942i	 &	0.8202-0.0945i &	 0.8806-0.0947i &	0.9593-0.0881i \\
\hline
5 \; (DO) & 1.0123 - 0.0949i & 1.0272 - 0.0953i & 1.1083 - 0.0955i & 1.2226 - 0.0875 \\
5 \; \text{\cite{refBHWKB3}} & 1.0123-0.0949i &	 1.027-0.0953i &	 1.1083-0.0955i &	1.2137-0.0884i \\
5 \; \text{\cite{refGunter1980RN}} & 1.0123-0.0949i &	 1.0272-0.0953i	& 1.1083-0.0955i &	1.2138-0.0887i \\
\hline
\end{tabular}
\label{table:RNGravLow}
\end{table}

\begin{table}[t]
\caption{\textit{Spin-1/2 QNFs for 4D RN BHs calculated using the DO expansion of Table \ref{table:DOschwarz} and compared with the PT results of Ref. \cite{refJing2003}.}}
\begin{tabular}{|C|C|C|C|C|}
\hline
\ell & \omega \; (\theta=0.0) & \omega \; (\theta=0.2) &  \omega \; (\theta=0.4) &\omega  \;(\theta =0.6)\\
\hline
1 \; (DO) & 0.3784 - 0.0930i & 0.3833 - 0.0930i & 0.3915 - 0.0936i & 0.4054 - 0.0954i \\
1 \; \text{\cite{refJing2003}} & 0.3855-0.0991i & 0.3881-0.0993i & 0.3964-0.0997i	& 0.4121-0.1003i \\
\hline
4 \; (DO) & 0.9603 - 0.0962i & 0.9673 - 0.0965i & 0.988 - 0.0971i & 1.027- 0.0978i \\
4 \; \text{\cite{refJing2003}} & 0.9625-0.0966i &	0.9690-0.0968i & 0.9898-0.0974i & 1.0293-0.0982i \\
\hline
\end{tabular}
\label{table:RNDiracLow}
\end{table}

\begin{table}[t]
\caption{\textit{Spin-3/2 QNFs for 4D RN BHs calculated using the DO method and compared with the  WKB and improved AIM results of Ref. \cite{refRSRN18}.}}
\begin{tabular}{|C|C|C|C|}
\hline
\ell & \omega \; (\theta =0.1) & \omega \; (\theta =0.5) &   \omega  \;(\theta =1.0)\\
\hline
0 \; (DO) & 0.3189 - 0.08516i & 0.3023 - 0.0506i & 0.5204 - 0.0766i\\
0 \; (WKB) & 0.3185-0.0910i  & 0.3634-0.0952i & 0.5414-0.0865i\\
0 \; (AIM) & 0.3185-0.0909i & 0.3621-0.0881i & 0.5414-0.0864i \\
\hline
1 \; (DO) & 0.5382 - 0.0932 & 0.5790 - 0.0944i & 0.8159 - 0.0870i \\
1 \; (WKB) & 0.5375-0.0942i & 0.5908-0.0971i & 0.8174-0.0874i\\
1\; (AIM) & 0.5375-0.0942i & 0.4050-0.1854i & 0.8174-0.0873i \\
\hline
2 \; (DO) & 0.7429 - 0.0950i &  0.7987 - 0.0970i & 1.0809 - 0.0879i \\
2 \; (WKB) & 0.7425-0.0952i & 0.8046-0.0976i & 1.0811-0.0878i\\
2 \; (AIM) & 0.7425-0.0952i & 0.8045-0.0975i & 1.0811-0.0877i \\
\hline
3 \; (DO) & 0.9427 - 0.09554i & 1.009 - 0.0973i & 1.3396 - 0.0881i \\
3 \; (WKB) & 0.9424-0.0957i & 1.0132-0.0977i & 1.3396-0.0880i\\
3 \; (AIM) & 0.9424-0.0957i & 1.0131-0.0977i & 1.3395-0.0879i \\
\hline
4 \; (DO) & 1.1400 - 0.0958i & 1.2166 - 0.0974i & 1.5953 - 0.0882i \\
4 \; (WKB) & 1.1399-0.0959i & 1.2193-0.0978i & 1.5953-0.0881i\\
4 \; (AIM) & 1.1399-0.0959i & 1.2193-0.0978i & 1.5953-0.0881i \\
\hline
5 \; (DO)& 1.3361 - 0.0960i & 1.4220 - 0.0975i & 1.8495 - 0.0882 \\
5 \; (WKB) & 1.3360-0.0960i & 1.4241-0.0978i & 1.8494-0.0882i\\
5 \; (AIM) & 1.3360-0.0960i & 1.4241-0.0978i & 1.8494-0.0882i \\
\hline
\end{tabular}
\label{table:RNRSLow}
\end{table}

\clearpage
\subsection{QNFs in Schwarzschild de Sitter BH spacetimes \label{app:SdS}}
\par In Ref. \cite{refZhidenko2004}, Zhidenko performed a numerical analysis on low-lying QNFs of spin $s \in \{0,1/2,1,2 \}$ within the SdS BH spacetime, using the 6th-order WKB and PT approximation methods. We use these results for our comparison. For the spin-3/2 QNFs, we utilise the notebook provided in Ref. \cite{refZhidenko2004} to compute the outstanding spin-3/2 results. 

\par For the SdS BH spacetimes, the DO method extends to order $\mathcal{O}(L^{-6})$ with relative ease. As such, all QNFs here are calculated to this order. We consistently find a closer agreement between the WKB and DO results than the WKB and the PT for all but the spin-3/2 QNFs, demonstrating that the DO method is highly reliable.

\begin{table}[b]
\caption{\textit{Spin-0 QNFs of 4D SdS BHs calculated using the DO expansions of Table \ref{table:DOSdS} and compared with the 6th-order WKB and PT results from Ref. \cite{refZhidenko2004}.}}
\begin{tabular}{|c|c|c|c|c|}
\hline
$\ell$ & $\omega$ $(\Lambda = 0.00)$ & $\omega$ $(\Lambda = 0.04)$ & $\omega$ $(\Lambda = 0.08)$ & $\omega$ $(\Lambda = 0.10)$ \\
\hline
$1$ ($DO$) & $0.2929-0.0976 i$ & $0.2246-0.0819 i$ & $0.1404-0.0540 i$ & $0.08160-0.03125 i$ \\
$1$ ($WKB$) & $0.2929-0.0978i$ & $0.2247-0.0821i$ & $0.1404-0.0542i$ & $0.08156-0.03124i$ \\
$1$ ($PT$) &$0.2990-0.1010i$ & $0.2260i-0.0830i$ & $0.1410-0.0550i$ &$0.0819-0.0315i$ \\
\hline
$2$ ($DO$) & $0.4836-0.0967 i$ &  $0.3808-0.0787 i$  & $0.2475-0.0519 i$ & $0.1466-0.0307 i$ \\
$2$ ($WKB$) & $0.4836-0.0968i$ & $0.3808-0.0788i$ & $ 0.2475-0.05197i$ & $0.1466-0.0307i$ \\
$2$ ($PT$) & $0.4870-0.0980i$ & $0.3820-0.0790i$ & $0.2480-0.0520i$ & $0.1468-0.0308i$ \\
\hline
\end{tabular}
\label{table:SdSscalarLow}
\end{table}

\begin{table}[b]
\caption{\textit{Spin-1 QNFs of 4D SdS BHs calculated using the DO expansions of Table \ref{table:DOSdS} and compared with the 6th-order WKB and PT results from Ref. \cite{refZhidenko2004}.}}
\begin{tabular}{|c|c|c|c|c|}
\hline
$\ell$ & $\omega$ $(\Lambda = 0.00)$ & $\omega$ $(\Lambda = 0.04)$ & $\omega$ $(\Lambda = 0.08)$ & $\omega$ $(\Lambda = 0.10)$ \\
\hline
$1$ ($DO$) & $0.2494-0.0921 i$ & $0.2010-0.0747 i$ & $0.1340-0.0502 i$ & $0.0804-0.0303 i$ \\
$1$ ($WKB$) & $0.2482-0.0926i$ & $0.2006-0.0748i$ &$0.1339-0.0502i$ &$0.0804-0.0303i$ \\
$1$ ($PT$) &$0.2550-0.0960i$ & $0.2040-0.0770i$  &$0.1350-0.0510i$ & $0.0805-0.0304i$ \\
\hline
$2$ ($DO$) & $0.4577-0.0950 i$ &  $0.3673-0.0762 i$  & $0.2437-0.0507 i$ & $0.1458-0.0304 i$ \\
$2$ ($WKB$) &$0.4576-0.09501i$ &$0.36723-0.07624i$ &$0.24365-0.0506i$ &$0.14582-0.03037i$ \\
$2$ ($PT$) &$0.4610-0.0960i$ &$0.3690-0.0770i$ & $0.2440-0.0510i$ & $0.1459-0.0304i$ \\
\hline
\end{tabular}
\label{table:SdSEMLow}
\end{table}

\begin{table}[t]
\caption{\textit{Spin-2 QNFs of 4D SdS BHs calculated using the DO expansions of Table \ref{table:DOSdS} and compared with the 6th-order WKB and PT results from Ref. \cite{refZhidenko2004}.}}
\begin{tabular}{|c|c|c|c|c|}
\hline
$\ell$ & $\omega$ $(\Lambda = 0.00)$ & $\omega$ $(\Lambda = 0.04)$ & $\omega$ $(\Lambda = 0.08)$ & $\omega$ $(\Lambda = 0.10)$ \\
\hline
$2$ ($DO$) & $0.3747-0.0808i$ & $0.2999-0.0702i$ & $0.1980-0.0493i$ & $0.1182-0.0301i$ \\
$2$ ($WKB$) &$0.3736-0.0889i$ & $0.2989-0.0733i$  &$0.1975-0.0499i$ &$0.11792-0.03021i$ \\
$2$ ($PT$) & $0.3780-0.0910i$ &$0.3010-0.0740i$ & $0.1980-0.0500i$ & $0.1180-0.0303i$ \\
\hline
$3$ ($DO$) & $0.5996-0.0911i$ & $0.4802-0.0745i$ & $0.3179-0.0503i$ & $0.1900-0.0303i$\\
$3$ ($WKB$) &$0.5994-0.0927i$ & $0.4801-0.07516$ & $0.3178-0.0504i$ &$0.1900-0.03031i$ \\
$3$ ($PT$) & $0.6020-0.0930i$ & $0.4810-0.0760i$ &$0.3180-0.0510i$ &$0.1901-0.0303i$ \\
\hline
\end{tabular}
\label{table:SdSGravLow}
\end{table}

\begin{table}[t]
\caption{\textit{Spin-1/2 QNFs of 4D SdS BHs calculated via the DO expansions of Table \ref{table:DOSdS} and compared with the 6th-order WKB and PT results from Ref. \cite{refZhidenko2004}.}}
\begin{tabular}{|c|c|c|c|c|}
\hline
$\ell$ & $\omega$ $(\Lambda = 0.00)$ & $\omega$ $(\Lambda = 0.04)$ & $\omega$ $(\Lambda = 0.08)$ & $\omega$ $(\Lambda = 0.10)$ \\
\hline
$0$ ($DO$) & $0.1810-0.0998i$ & $0.1473-0.0768i$ & $0.1003-0.0505i$ & $0.0606-0.0303i$ \\
$0$ ($WKB$) &$0.1830-0.0950i$ & $0.1490-0.0750i$ & $0.1006-0.0501i$ & $0.0606-0.0303i$ \\
$0$ ($PT$) &$0.1890-0.1050i$& $0.1510-0.0820i$ &$0.1001-0.0529i$ &$0.0597-0.0311i$\\
\hline
$1$ ($DO$) & $0.3798-0.0965i$ & $0.3053-0.0769i$ & $0.2029-0.0508i$ & $0.1216-0.0304i$\\
$1$ ($WKB$) &$0.3801-0.0964i$ &$0.3054-0.0769i$ &$0.20296-0.0508i$ &$0.1216-0.0304i$ \\
$1$ ($PT$) &$0.3860-0.0990i$ &$0.3080-0.0790i$ &$0.2040-0.0520i$ &$0.1220-0.0305i$ \\
\hline
\end{tabular}
\label{table:SdSDiracLow}
\end{table}

\begin{table}[t]
\caption{\textit{Spin-3/2 QNFs of 4D SdS BHs calculated using the DO expansions of Table \ref{table:DOSdS} and compared with the 6th-order WKB and PT results from Ref. \cite{refZhidenko2004}.}}
\begin{tabular}{|c|c|c|c|c|}
\hline
$\ell$ & $\omega$ $(\Lambda = 0.00)$ & $\omega$ $(\Lambda = 0.04)$ & $\omega$ $(\Lambda = 0.08)$ & $\omega$ $(\Lambda = 0.10)$ \\
\hline
$2$ ($DO$) & $0.7349-0.0948i$ & $0.5876-0.0764i$ & $0.3882-0.0509i$ & $0.2318-0.0305i$\\
$2$ ($WKB$) &$0.7348-0.0949i$ & $0.5885-0.0763i$ & $0.3897-0.0507i$ & $0.2348-0.0302i$ \\
$2$ ($PT$) & $0.7366-0.0952i$ & $0.5893-0.0764i$ & $0.3898-0.0507i$ & $0.2330-0.03049$ \\
\hline
$3$ ($DO$) & $0.9332-0.09534i$ & $0.7465-0.0766i$ & $0.4936-0.0509i$ & $0.2949-0.0304i$\\
$3$ ($WKB$) & $0.9344-0.0954i$ & $0.7480-0.0765i$ & $0.4951-0.0509i$ & $0.2964-0.0304i$\\
$3$ ($PT$) & $0.9359-0.0956i$  & $0.7488-0.0766i$ & $0.4953-0.0508i$ & $0.2960-0.0304i$\\
\hline
\end{tabular}
\label{table:SdSRSLow}
\end{table}
\vfill

\clearpage
\bibliographystyle{aipnum4-2} 
\bibliography{bibexport}

\end{document}